\documentclass{aastex631} %
\usepackage{amsmath}
\usepackage{gensymb}

\shorttitle{Thermal Models of Asteroids with Two-band Combinations of WISE Cryogenic Data}
\shortauthors{Whittaker, Margot, Lam, Myhrvold}

\usepackage{hyperref}
\hypersetup{colorlinks=true,citecolor=blue,linkcolor=blue,urlcolor=blue,breaklinks=true}

\begin{document}
\title{Thermal Models of Asteroids with Two-band Combinations of Wide-field Infrared Survey Explorer Cryogenic Data}

\correspondingauthor{emilywhittaker@g.ucla.edu}

\author[0000-0002-1518-7475]{Emily A. Whittaker}
\affiliation{Department of Earth, Planetary, and Space Sciences, University of California, Los Angeles, CA 90095, USA}

\author[0000-0001-9798-1797]{Jean-Luc Margot}
\affiliation{Department of Earth, Planetary, and Space Sciences, University of California, Los Angeles, CA 90095, USA}
\affiliation{Department of Physics and Astronomy, University of California, Los Angeles, CA 90095, USA}

\author[0000-0002-4688-314X]{Adrian L.~H. Lam}
\affiliation{Department of Electrical Engineering, University of California, Los Angeles, CA 90095, USA}

\author[0000-0003-3994-5143]{Nathan Myhrvold}
\affiliation{Intellectual Ventures, 3150 139th Ave SE, Bellevue, WA 98005, USA}

\begin{abstract}

We used the reparameterized Near-Earth Asteroid Thermal Model to model observations of a curated set of over
4000 asteroids from the Wide-field Infrared Survey Explorer in two wavelength bands (W2--3 or W3--4) and
compared the results to previous results from all four wavelength bands (W1--4). This comparison was done with
the goal of elucidating unique aspects of modeling two-band observations so that any potential biases or
shortcomings for planned two-band surveys (e.g., the NASA Near-Earth Object Surveyor Mission) can be
anticipated and quantified. The W2--3 two-band fits usually yielded slightly smaller diameters than the four-band
fits, with a median diameter difference of -10\%, with the 5\% and 95\% quantiles of the distribution at -32\% and
-1.5\%, respectively. We conducted similar comparisons for W3--4, in part because the longest wavelength bands
are expected to provide the best two-band results. We found that the W3--4 two-band diameters are slightly larger
than the four-band results, with a median diameter difference of 11\% and the 5\% and 95\% quantiles of the
distribution at -2.1\% and 26\%, respectively. The diameter uncertainty, obtained with bootstrap analysis, is larger
by 30\% and 35\% (median values) for the W2--3 and W3--4 fits, respectively, than for the corresponding four-band
fits. Using 23 high-quality stellar occultation diameters as a benchmark, we found that the median errors of W2--3
and W3--4 diameter estimates are -15\% and +12\%, respectively, whereas the median error of the four-band fits is
9.3\%. Although the W2--3 and W3--4 diameters appear to have greater systematic errors and uncertainties than
their four-band counterparts, two-band estimates remain useful because they improve upon diameter estimates
obtained from visible photometry alone.
\end{abstract}

\section{Introduction}
\label{sec-intro}

The Wide Field Infrared Survey Explorer (WISE) is a space-based telescope that conducted a full-sky survey in the infrared in four wavelength bands 
centered at 3.4, 4.6, 12, and 22 $\mu$m (called W1, W2, W3, and W4, respectively).  This survey yielded observations of over 100,000 asteroids and the 
opportunity to provide estimates of size, temperature, albedo, and sometimes even rotation period for many asteroids \citep{wrig10}. 

The massive task of analyzing these asteroid observations was pioneered by the NEOWISE team, who used the Near-Earth Asteroid Thermal Model (NEATM)
(\citet{Harris_1998}, but see also \citet{myhr18})
to fit for a variety of parameters for each asteroid such as diameter, temperature, and albedo depending on the number, quality, and band distribution of the flux measurements available for each object.  A curated subset of these observations were reanalyzed by \citet{myhr22}, who used a reparameterization of the 
NEATM to eliminate unnecessary degrees of freedom within the model, avoid the assumption of constant emissivities, improve the goodness-of-fit to observables, and obtain more accurate parameter estimates.

Of the asteroid observations analyzed by the NEOWISE team, asteroids with observations in only W3 and W4 outnumber asteroids with observations in all four bands by 19:1 \citep{myhr18empirical}.  Additionally, of all possible two-band combinations, W3 and W4 are expected to give the best results because these wavelengths are
near the peak of the spectral energy distribution for asteroid thermal emission and experience less contamination from reflected sunlight, which peaks in the visible range.  Therefore, it is important to compare the W3--4 results with the four-band results to quantify any uncertainties and potential biases for this large data set.

The anticipated NASA Near-Earth Object Surveyor Mission (NEO Surveyor) is a 50 cm telescope designed to expand the inventory of asteroids, including the identification of potentially hazardous asteroids (PHAs).
Unlike WISE, NEO Surveyor may operate with two wavelength bands (4--5.2 $\mu$m and 6--10 $\mu$m) \citep{Mainzer_2016b}, which correspond roughly to WISE bands W2 and W3.  While W2 and W3 do not span exactly the same wavelengths as the anticipated NEO Surveyor bands, they are similar enough that they can provide an estimate of what to expect from NEO Surveyor's future observations.
Our sample of data from the WISE cryogenic mission includes observation of 4420 asteroids.  Approximately 5\% of the asteroids in this sample were observed in two distinct observational sequences separated by more than 30 days.  Separate solutions were calculated for the two clusters.  Our sample includes 4685 clusters of observations.

In this work, 
we conducted two sets of two-band (W2--3 and W3--4) thermal fits of the 4685 clusters of asteroid observations analyzed with four bands by \citet{myhr22} to quantify the uncertainties and potential biases associated with two-band fits.
Understanding the strengths and weaknesses of two-band instruments in estimating asteroid parameters may assist the planetary defense community in predicting what new information NEO Surveyor is likely to provide about the PHA population, while also giving us the opportunity to anticipate weaknesses and blind spots that may hinder a PHA’s characterization.

\section{Methods}
\label{sec-methods}

\subsection{Data Selection}
In this work, we used the same dataset that was used by \citet{myhr22}, but we focused on two wavelength bands,
either W2--3 or W3--4, instead of all four infrared bands.  We provide a brief description of how this data set was selected; for details, see \citet{myhr22}.

Photometric data were obtained from the WISE full cryogenic mission phase, between January 7 and August 6 of 2010.  First, a ``moving object search"
was performed on the WISE All-Sky Single Exposure Level 1b Source Catalog for all asteroids in the Minor Planet Center Orbit Database, returning a
list of observation dates as Modified Julian Date (MJD) values \citep{All-sky20}.  The JPL HORIZONS service was used to obtain the relevant parameters for each asteroid
on the given observation dates including H and G values, positions, distances to the Sun and to the WISE spacecraft, and Sun-Target-Observer (phase) 
angles.  This step yielded about 3.2 million records.  A catalog cone search with a radius of 10 arcsec was performed around each of these positions, and all detected sources were stored for future processing.  

After this initial selection, multiple filters were applied to the data.  First, we removed asteroid conjunctions, where two or more presumed-asteroid sources were identified within 10 arcsec of another.  A total of 12,000 conjunctions were found.  

Next, we removed 419,039  observations with low ($<$2) signal-to-noise ratio in W3 or W4, where asteroids are typically brightest.

A further 96,685 observations were eliminated on the basis of five quality indicators reported within the WISE database, which were applied on a
per-band basis.  Observations were discarded if they had nonexistent (``Null'') or poor (``ph\_qual'' $\neq$ ``A'' or ``B'') flux determinations, if they were susceptible to contamination or confusion (``cc\_flags'' $\neq$ 0), if they had a low signal-to-noise ratio, (``w\textit{k}snr'' $<$ 4), if they were saturated (``w\textit{k}sat'' $\neq$ 0), or if they had poor point-spread-function fits (``w\textit{k}rchi2'' $>$ 2) ($1\leq k \leq4$).

The fourth filter was designed to eliminate potential confusion between asteroids and other astronomical sources listed in the AllWISE
(co-added) Source Catalog.  Every AllWISE source within 10 arcsec of an ephemeris position from one of the remaining potential asteroids was listed as a 
potentially confusing background source and further examined.  This step eliminated approximately 13,000 observations.

The fifth filter was based on limiting the allowable distance between the WISE observed position and the predicted ephemeris position to specific thresholds \citep[][Table 1]{myhr22}.  

The sixth filter split the observations for each asteroid into distinct clusters of observations if consecutive observations were more than 30 days apart.  
Following NEOWISE practice, we determined independent solutions for each cluster of observations.
94\% of asteroids were observed in a single cluster.
Any cluster that did not contain at least three observations in each of the four WISE wavelength bands was removed.  

Lastly, any observation $\geq$ 1.5 mag away from all other fluxes within its cluster was removed.  

After filtering, we were left with 82,548 observations grouped into 4685 clusters, representing 4420 asteroids.

\subsection{NEATM Reparameterization}
\label{sec-reparam} 
The original derivation of the NEATM hypothesizes a temperature distribution across the asteroid according to the following expression 
\citep{Harris_1998, Harris_and_Lagerros_2002}:
\begin{equation}
T_{\rm NEATM}(\theta, \phi, r_{\rm as}) = T_{\rm ss}(r_{\rm as})(\sin \theta \cos \phi)^{0.25},
\label{eqn:NEATM temp dist}
\end{equation}
where $\theta$ and $\phi$ are the zenithal and azimuthal angles in a spherical coordinate system and $r_{\rm as}$ is the asteroid-Sun distance
in astronomical units (au).  In NEATM, the maximum 
temperature across the surface is the subsolar temperature:
\begin{equation}
  T_{\rm ss}(r_{\rm as}) = \left(\frac{S(1-p_{v}q)}{\epsilon_{B}\sigma\eta r_{\rm as}^2}\right)^{0.25},
\label{eqn:NEATM temp ss}
\end{equation}
where $S$ is the solar constant (1360.8 W/m$^2$), $p_v$ is the visible geometric albedo, $q$ is the phase integral, $\epsilon_B$ is the emissivity, $\sigma$ is the 
Stefan-Boltzmann constant, and $\eta$ is the beaming parameter, defined as ``the inverse of the enhancement in total thermal flux at zero solar 
phase angle over that of a uniformly radiating sphere'' \citep{Lebofsky_1986}.
See \citet[][]{myhr18} for an extensive discussion of the NEATM model derivation, assumptions, and limitations.

Results obtained by the NEOWISE team and archived at the PDS \citep{Mainzer_2016a} were obtained with 10 different combinations
of
adjustable parameters.
For a list of model names used by the NEOWISE team, see Table 1 of \citet{myhr18empirical}.  The NEOWISE team assumed that the emissivity is 0.9 in each wavelength band.

Many studies arbitrarily set $\epsilon=0.9$ \citep{delbo04}.  In NEOWISE work, the emissivity is held at 0.9 while the albedo is allowed to vary.
However, holding the emissivity constant while allowing the albedo to vary violates Kirchhoff’s 
law, which may introduce systematic errors into the model \citep{myhr18}.  
For an estimate of these errors, see Appendix \ref{app-4vs4pt9}.

In the presence of reflected sunlight, the flux equation involves both emitted and reflected light components \citep{myhr18}:
\begin{equation}
F_{\rm obs}(\lambda, \alpha, r_{\rm as}, r_{\rm ao}) = \frac{D^2}{4 {\rm au}^2 r_{\rm ao}^2} \times \left(\epsilon(\lambda) F_{\rm model}(\alpha, \lambda) + p(\lambda)\frac{\psi_{\rm HG}(\alpha, G)}{r_{\rm as}^2} F_{\rm Sun}(\lambda)\right), 
\label{eqn:F reflected Sun}
\end{equation}
where $\lambda$ is the wavelength, $\alpha$ is the phase angle, $r_{\rm ao}$ is the asteroid-observer distance in 
au, $\psi_{\rm HG}$ is the phase function, 
$F_{\rm model}(\alpha, \lambda)$ is the emitted flux specific to the adopted asteroid thermal model, and $F_{\rm Sun}(\lambda)$ is the solar flux incident on an asteroid at 1 au from the Sun, with $S=\int_0^\infty F_{\rm sun}(\lambda) d\lambda$.  %
In our formulation, the geometric albedo $p(\lambda)=(1 - \epsilon(\lambda))/q$ obeys Kirchoff's Law.

Because $F_{\rm model}$ for the NEATM depends on integrating the Planck distribution over all visible angles for an object, and the Planck distribution depends on $T_{\rm NEATM}$, it is important to consider how this temperature is included in the modeling process.  Using equation \ref{eqn:NEATM temp ss} in the thermal modeling process, as used by the NEOWISE team in \citet{masiero2011}, allows for the visible albedo $p_{v}$ and the beaming parameter $\eta$ to vary.  While the beaming parameter can be conceptually useful, it can cause problems in a least-squares fit that simultaneously solves for $\eta$ and the visible albedo $p_v$ because $\eta$ and $p_v$ appear in the numerator and the denominator of equation (\ref{eqn:NEATM temp ss}), respectively.  
One can avoid these problems with a reparameterization by introducing a pseudo-temperature %
\begin{equation}
T_{1} = T_{\rm ss}\sqrt{r_{\rm as}},
\label{eqn:T1}
\end{equation}
which represents the subsolar temperature of an asteroid with $r_{\rm as}=1$ au \citep{myhr18}.  
By instead fitting for $T_{1}$ in the thermal model, this reparameterization eliminates two parameters from the model,  $\eta$ and $p_v$, and restores the NEATM model (without reflected sunlight) to
the two-parameter model that it should be ($D$ and $T_1$ for fixed $\epsilon$), thus improving the chance that numerical fitting algorithms will 
retrieve an accurate result.

\citet{myhr22} showed that the assumption of a constant emissivity ($\epsilon=0.9$) across all four WISE bands is likely responsible for the fact that up to 50\% of NEOWISE model fits miss the data completely in one or more WISE bands.  For this reason, they modified the fitting procedure and allowed the emissivity $\epsilon$ to vary between 0.5 and 1 for all four wavelength bands.  However, they used a regularization term to penalize models that yielded a large dispersion in emissivities or emissivities that differed strongly from the canonical value of 0.9.  For details about this regularization term, see \citet[][Section 3.6]{myhr22}.  In their four-band fits, the 
adjustable parameters were $(D, T_1, \epsilon_1, \epsilon_2, \epsilon_3, \epsilon_4)$,
where the emissivity subscripts represent the WISE bands.

\subsection{Minimization Method}

A smaller number of parameters is required to fit two-band data.  The diameter and pseudo-temperature are sufficient to model the data, and the introduction of a floating emissivity is superfluous, as demonstrated by an analysis of variance (Appendix \ref{app-eps}).  Therefore, we set $\epsilon=0.9$ for both wavelength bands and fit for $D$ and $T_1$.
Because the regularization term is zero when $\epsilon=0.9$, the minimization term used in this work is simply the sum of squared residuals (SSR):
\begin{equation}
{\rm SSR} = \sum_{i=1}^{N}(x_{i, \rm model} - x_{i, \rm obs})^{2}
\label{eqn:gof}
\end{equation}
where $x_{i,\rm model} $ are the modeled photometric values, $x_{i,\rm obs}$ are the observed photometric values, and $N$ is the number of flux measurements for an object.
Following \citet{myhr22}, we chose to discard the flux uncertainties from the calculation of this minimization term because they are similar across flux measurements and can be much smaller than light curve variations due to the asteroid's rotation.  For example, the median flux uncertainty of all median fluxes for each cluster in the W2--3 dataset is 0.14 mag while the median light curve amplitude for the asteroids with an associated calculated light curve (35\% of objects) is 0.50 mag. 
For an examination of the impact of lightcurve amplitude on the goodness-of-fit metric, see Appendix~\ref{app-chisq}.

Because the minimization methods are different between the two-band models with fixed emissivity and the four-band models with variable emissivity, we sought to verify that differences in parameter estimates were due to the number of bands used in the fits rather than the minimization method.  To that end, we also ran a set of fits where we fixed the emissivity at 0.9 and used the SSR loss term in the four-band fits, replicating the two-band fitting procedure exactly.  The results strongly suggest that the observed differences between two-band and four-band fits is due to the consequences of eliminating two bands from the fits rather than any difference between fitting methods (Appendix \ref{app-2vs4}).

\subsection{Bootstrap Trials for Uncertainty Estimation}
\begin{figure}
\centering\includegraphics[width=7cm]{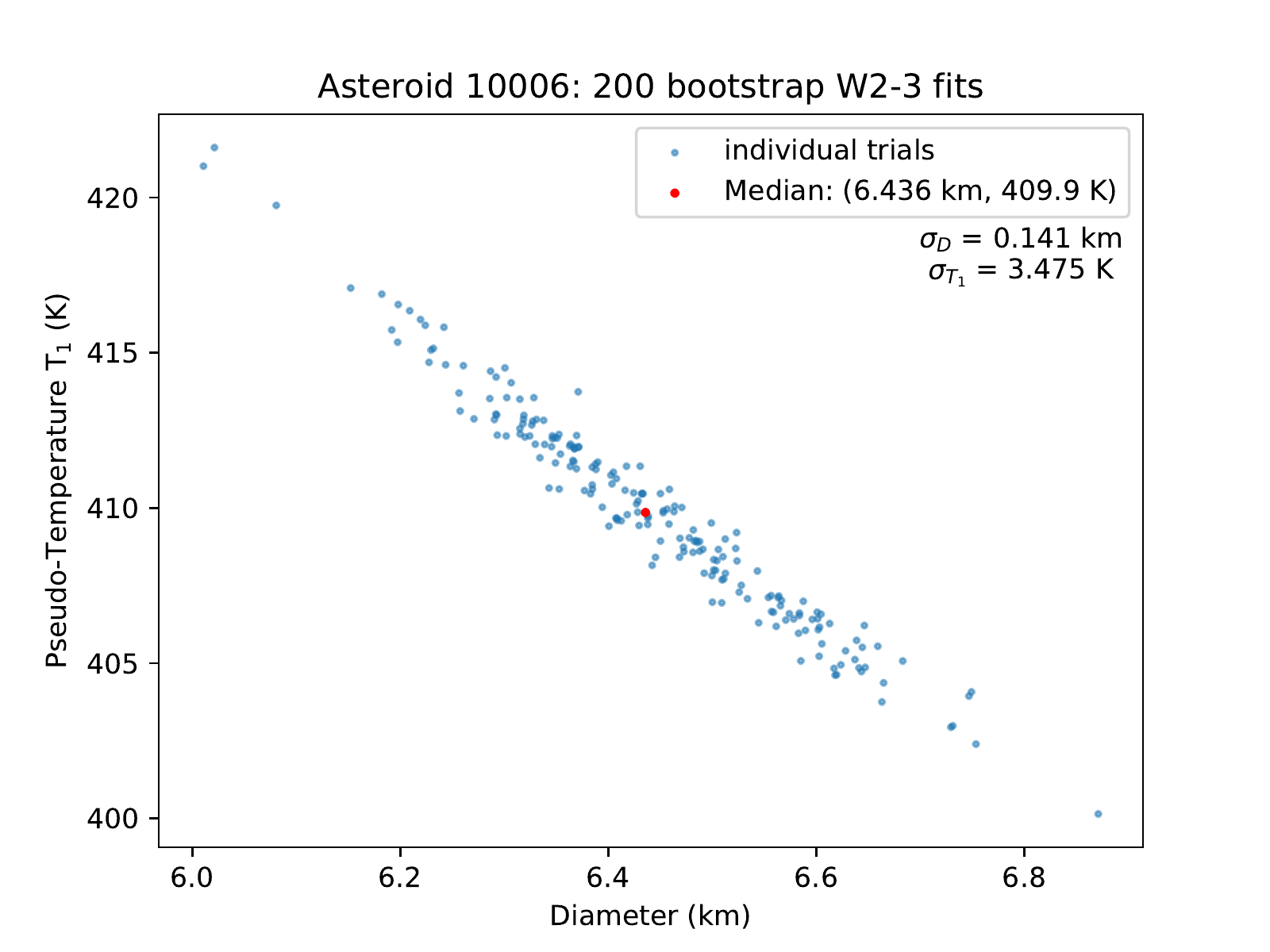}
\centering\includegraphics[width=7cm]{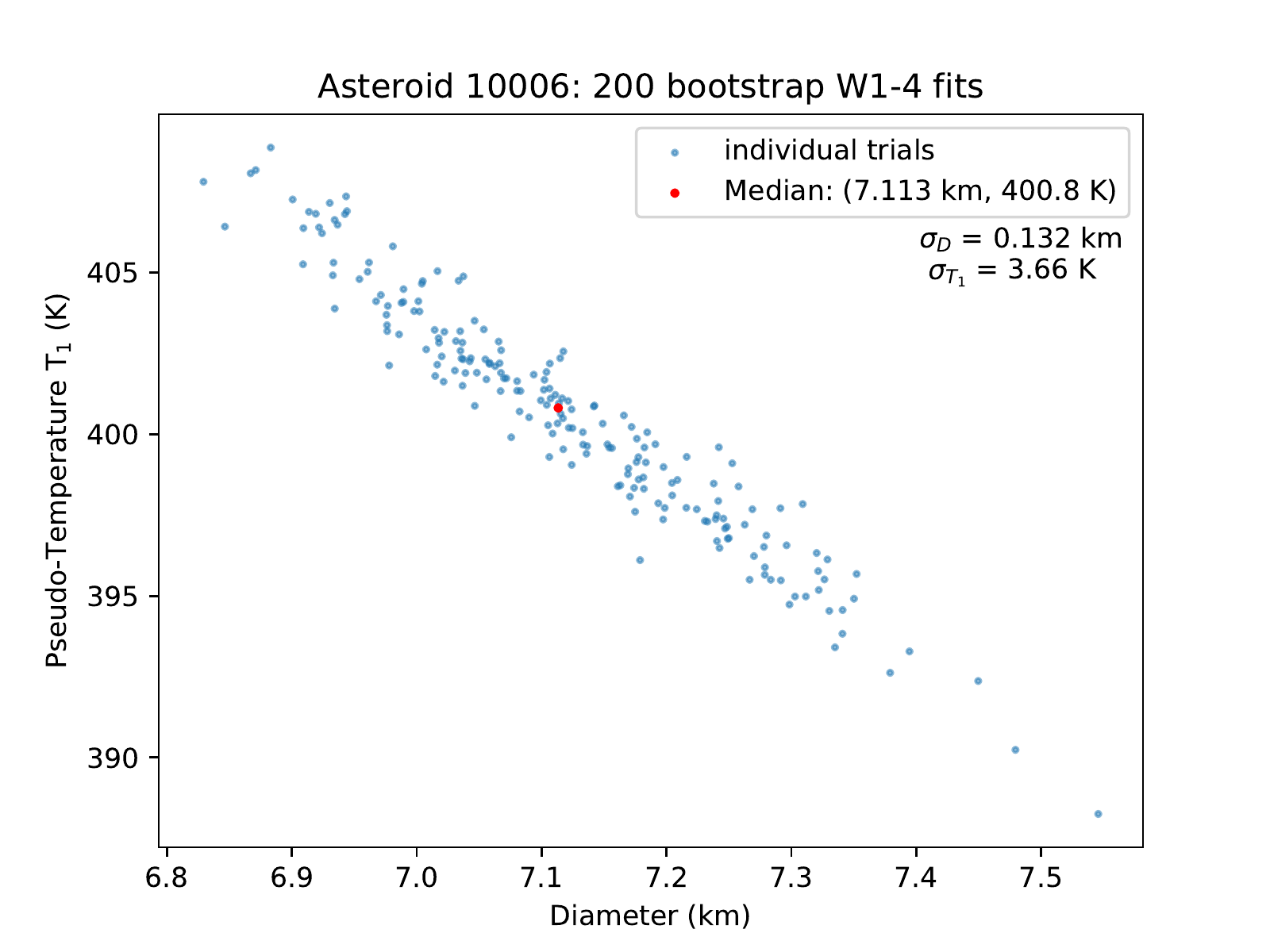}
\caption{Best-fit pseudo-temperature and diameter for 200 bootstrap trials for example asteroid 10006.  The median is shown as a red dot, with 
corresponding parameter values displayed in the legend.  Standard deviations in diameter and pseudo-temperature amongst the bootstrap trials
are displayed below the legend.  Left: Results from two-band modeling.  Right: Results from four-band modeling, where we replicate the results from \citet{myhr22} with the same set of observational clusters but a different set of random bootstrap trials.}
\label{fig:10006_bootstraps}
\end{figure}

We used a bootstrap method to obtain estimates of the parameter uncertainties.
We performed 200 bootstrap trials for each cluster, where, for each trial, we randomly sampled the asteroid flux measurements with replacement such that the number 
of data points was the same as it was originally.  If this resampling resulted in fewer than 3 datapoints in
either band, then we repeated the random sampling until there were at least 3 datapoints in each band included in the fit.  A model was fit to each resampled data set, resulting in 200 solutions per cluster.
The final model parameters were obtained by taking the median of the 200 parameter estimates, and the parameter uncertainties were obtained from the standard deviations of the parameter values within the 200 trials.  An example of the bootstrap procedure with 200 model fits is shown in Figure \ref{fig:10006_bootstraps} for asteroid 10006, which was observed in only one cluster.

For a random subset of 50 asteroids, we expanded the number of trials to 2000 and compared the 2000- and 200-trial results.  We found that median values were essentially unchanged, whereas estimates of uncertainties differed by up to 10\%.

\subsection{Treatment of Low Pseudo-Temperature Solutions}
\begin{figure}
  \centering\includegraphics[width=10cm]{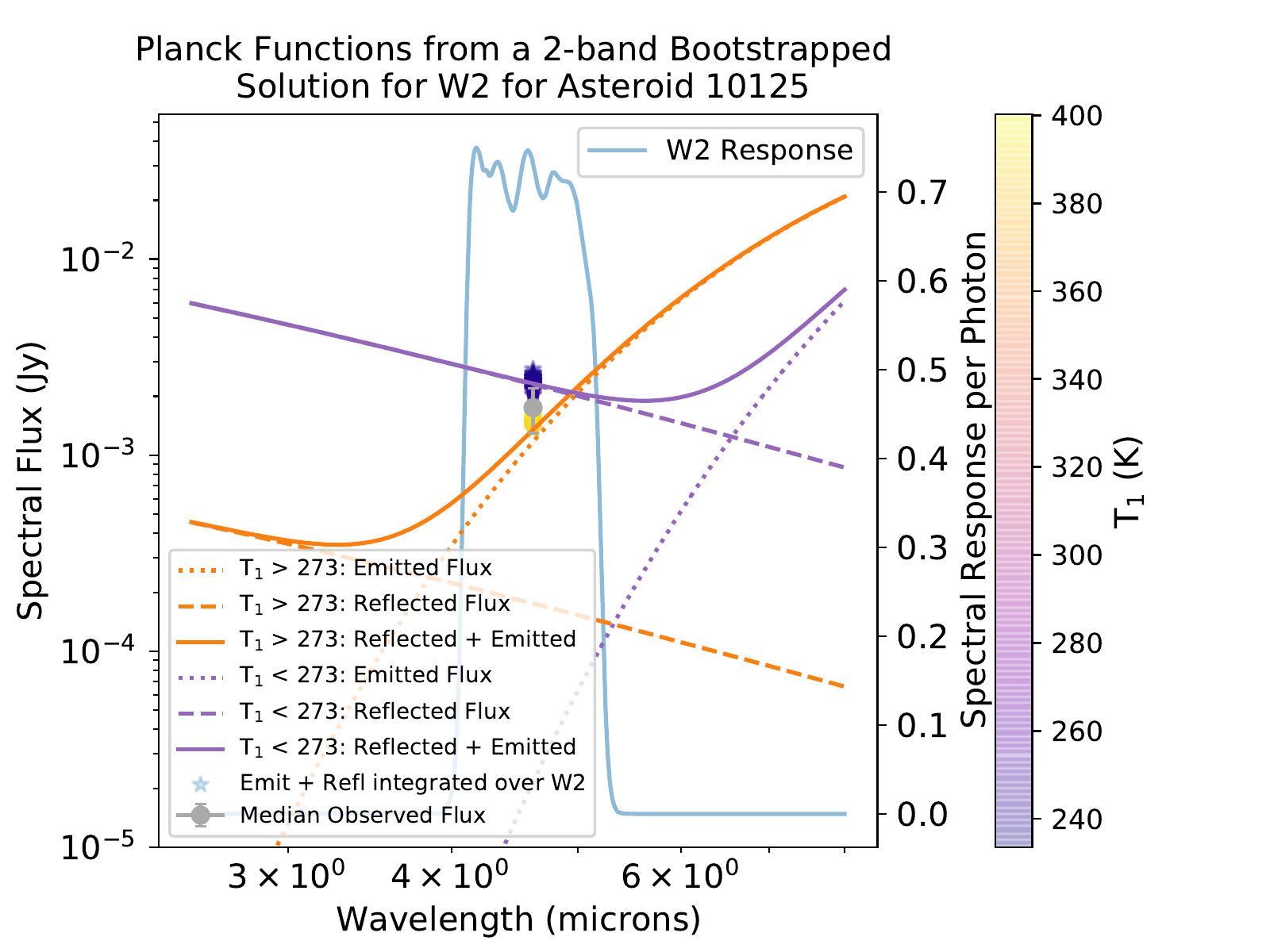}
  \\
  \centering\includegraphics[width=7cm]{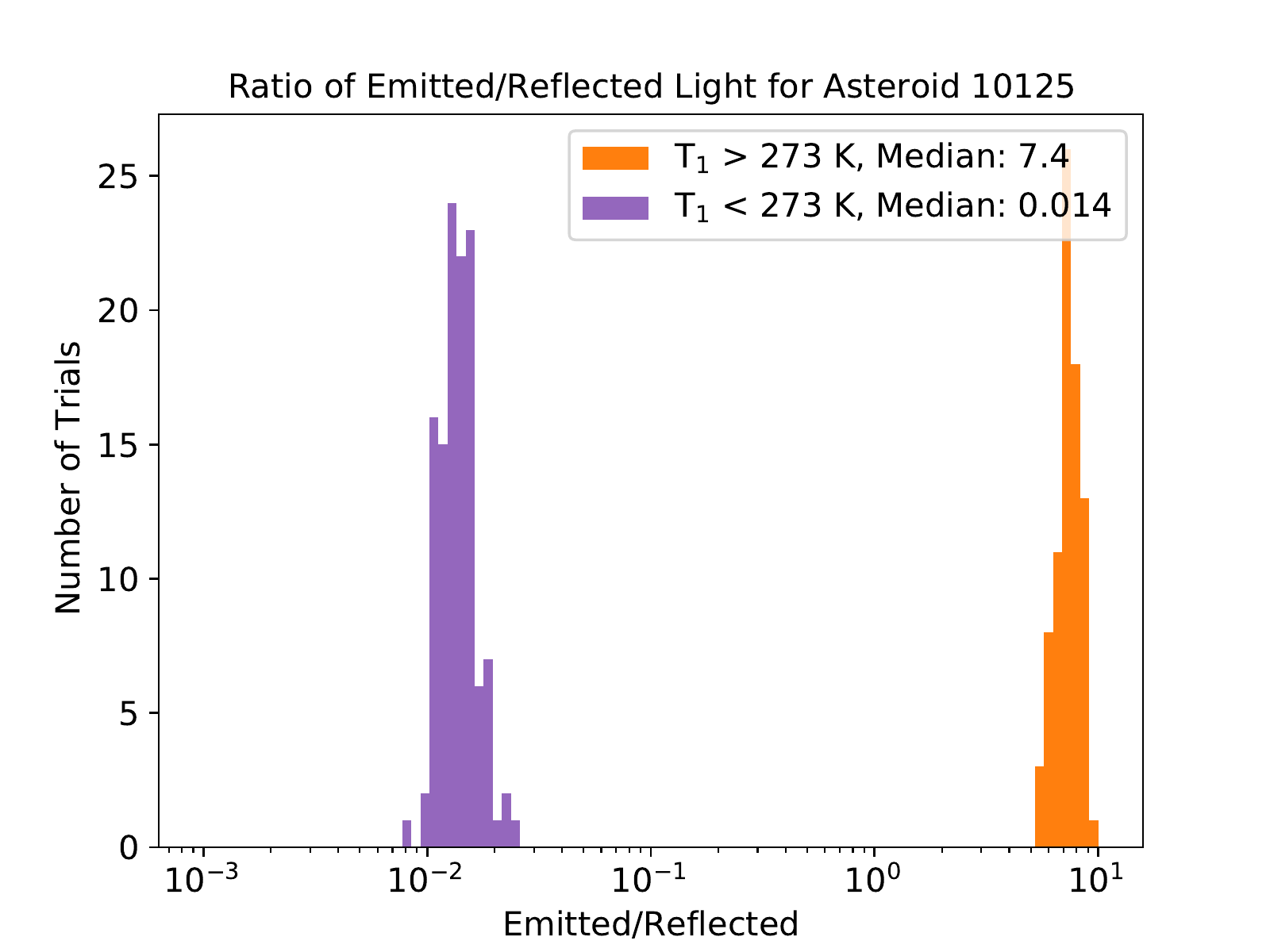}
  \centering\includegraphics[width=7cm]{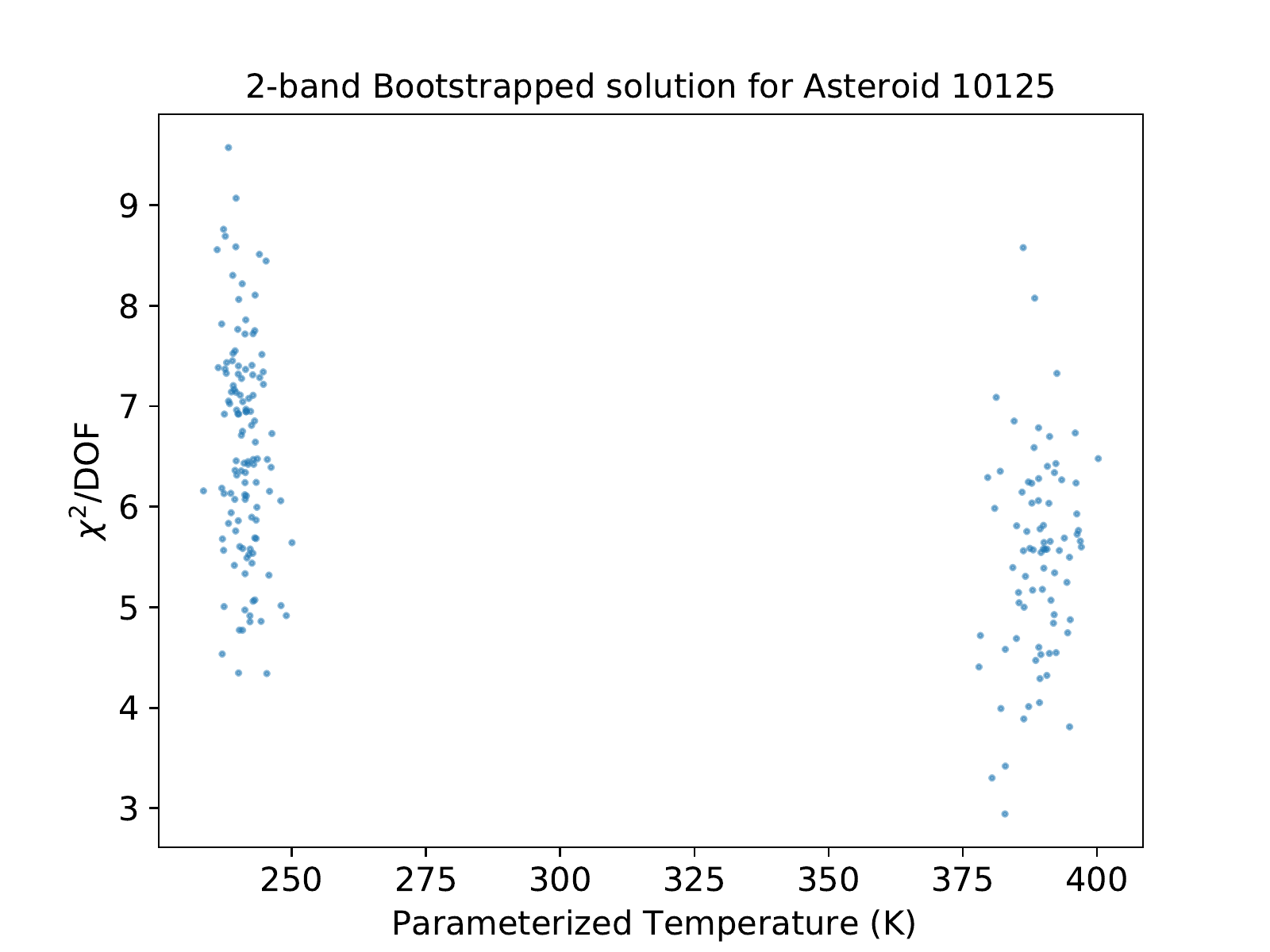}
  \caption{Observed dichotomy in pseudo-temperature solutions for two-band fits illustrated with results from 200 bootstrap trials for asteroid 10125.  Top: Planck spectral distributions of emitted light (dotted lines) and reflected light (dashed lines) calculated with the median pseudo-temperature $T_1$ and median diameter $D$ obtained from
the bootstrap trials.  The orange curves correspond to the median of 80 trials with high pseudo-temperatures and the purple curves correspond to the median of 120 trials with low pseudo-temperatures.    
    Note how two dissimilar pseudo-temperatures yield a similar total flux (solid lines) within the W2 band, while also yielding similar W3 fluxes.
    Colored symbols indicate the total (emitted + reflected) spectral flux integrated over the W2 bandpass
    for each of the 200 trials.  The W2 bandpass is shown in a light blue solid line as the spectral response per photon \citep{wrig10}.  The grey circle represents the true median measured W2 flux of the object adjusted to a phase angle of zero, with the error bar given by the standard deviation of the phase-adjusted W2 fluxes. Lower Left: Histogram of emitted to reflected light (integrated over the W2 bandpass) for 200 bootstrapped trials.  High
    pseudo-temperature solutions are shown in orange whereas low
    pseudo-temperature solutions are shown in purple.  Lower Right: Reduced chi-squared vs.\ pseudo-temperature $T_1$, showing similar values in the two temperature regimes.}
  \label{fig:planck}
\end{figure}

For individual bootstrap trials of both W2--3 and W3--4, the modeled solutions occasionally yielded an unrealistically low 
pseudo-temperature, with $T_1$ $<$ 273~K.
Most solutions yielded more plausible pseudo-temperatures, with $T_1$ $>$ 273~K. Specifically, 
the median pseudo-temperature of the 200 bootstrapped trials for the two-band W2--3 model was $<$ 273~K for 5.4\% of the 4685 cases.  For the W3--4 model, 
all median pseudo-temperatures
were $>$ 273~K.
In each instance among the 5.4\% of anomalous W2--3 cases,
we observed a dichotomy of solutions, with some of the trials yielding a high pseudo-temperature and small diameter and other trials yielding a low pseudo-temperature and large diameter (Figure \ref{fig:planck}, Lower Right).

We suspect that the dichotomy in pseudo-temperatures is related to the fact that it is possible to achieve approximately the same
W2 flux
with vastly different contributions of emitted and reflected light (Figure \ref{fig:planck}, Top), while also maintaining approximately the same W3 flux.  For instance, the low-temperature fits (median ratio of integrated emitted to reflected light of 0.014) for asteroid 10125 tend to have comparable chi-squared values to the high-temperature fits (median ratio of integrated emitted to reflected light of 7.4) (Figure \ref{fig:planck}, Lower Left and Lower Right).  With noisy data, the fitting algorithm will occasionally land on an unphysical low-temperature solution because
its chi-squared value is marginally lower than that of the high-temperature solution.
These anomalous cases tended to have smaller heliocentric distances and larger phase angles than most objects in our sample.   These trends are further explored in Appendix \ref{app-lowT.}

In the case of the W3--4 models, there were no asteroids with
pseudo-temperatures below 273~K, but there were 18 clusters with pseudo-temperatures between 273~K and 300~K.  These clusters do not present a dichotomy of solutions in the bootstrap trials, instead having all pseudo-temperatures clustered within the same pseudo-temperature range.  The median distance to the Sun for these asteroids was 1.06 au, as opposed to the median for all clusters of 2.37 au,
suggesting that NEOs were again implicated.  
In fact, 17 of the 18 cases
are indeed NEOs, representing
$\sim$20\% of all NEOs in our dataset.  NEOs may be more susceptible to
inaccurate fits due to their higher observation phase angles, which
exacerbate NEATM limitations and
suboptimal phase-curve corrections.
This hypothesis is supported by
a median phase angle of 70$^\circ$ among the 18 low pseudo-temperature cases, in contrast to the median of 25$^\circ$ among all clusters.

It was important to detect and remove the anomalous low-temperature solutions.  For the four-band models, each case was modeled with seven different starting values of emissivity, which provided a natural way of discarding low-temperature solutions.  The maximum $T_1$ was identified as $T_{1, \rm max}$, and any solution with $T_1 < 0.95 \times T_{1, \rm max}$ was discarded.  For the two-band models where emissivity was not varied, a different approach had to be used.
If fewer than 10\% of the bootstrap trials yielded a high-temperature ($T_1$ $>$ 273 K) solution, the entire cluster was thrown out.
Otherwise, the trials with $T_1$ $<$ 273~K were thrown out.
For the W2--3 results, this process eliminated 3.05\% of two-band cases, leaving 4542 cases.  For the W3--4 results, this process eliminated no two-band cases.
We focused on these 4542 W2--W3 cases and 4685 W3--4 cases for the remainder of the analysis, where each case was compared to its four-band counterpart from \citet{myhr22}.

\subsection{Albedo Calculation}
The visible-band geometric albedo was calculated for each observation cluster according to the equation
\begin{equation}
p_v \equiv \left( \frac{1329}{D} \right)^2 10^{\frac{-2H}{5}},
\label{eqn:albedo}
\end{equation}
where $D$ is the median
diameter estimate from the bootstrap trials and $H$ is the absolute magnitude of the asteroid.
We used the $H$ values from \citet{veres15} if available\footnote{Except for asteroid 19761, whose H value was off by $\sim$5 in the 2015 table.}.  Otherwise, we used the $H$ value published by the Minor Planet Center/HORIZONS plus 0.26 to account for the mean systematic offset reported in  \citet{veres15}.  For further details, see \citet[][Section 4.2]{myhr22}.

\section{Results}
\label{sec-results}

\subsection{W2--3 Results}

\begin{figure} 
\centering\includegraphics[width=.6\linewidth]{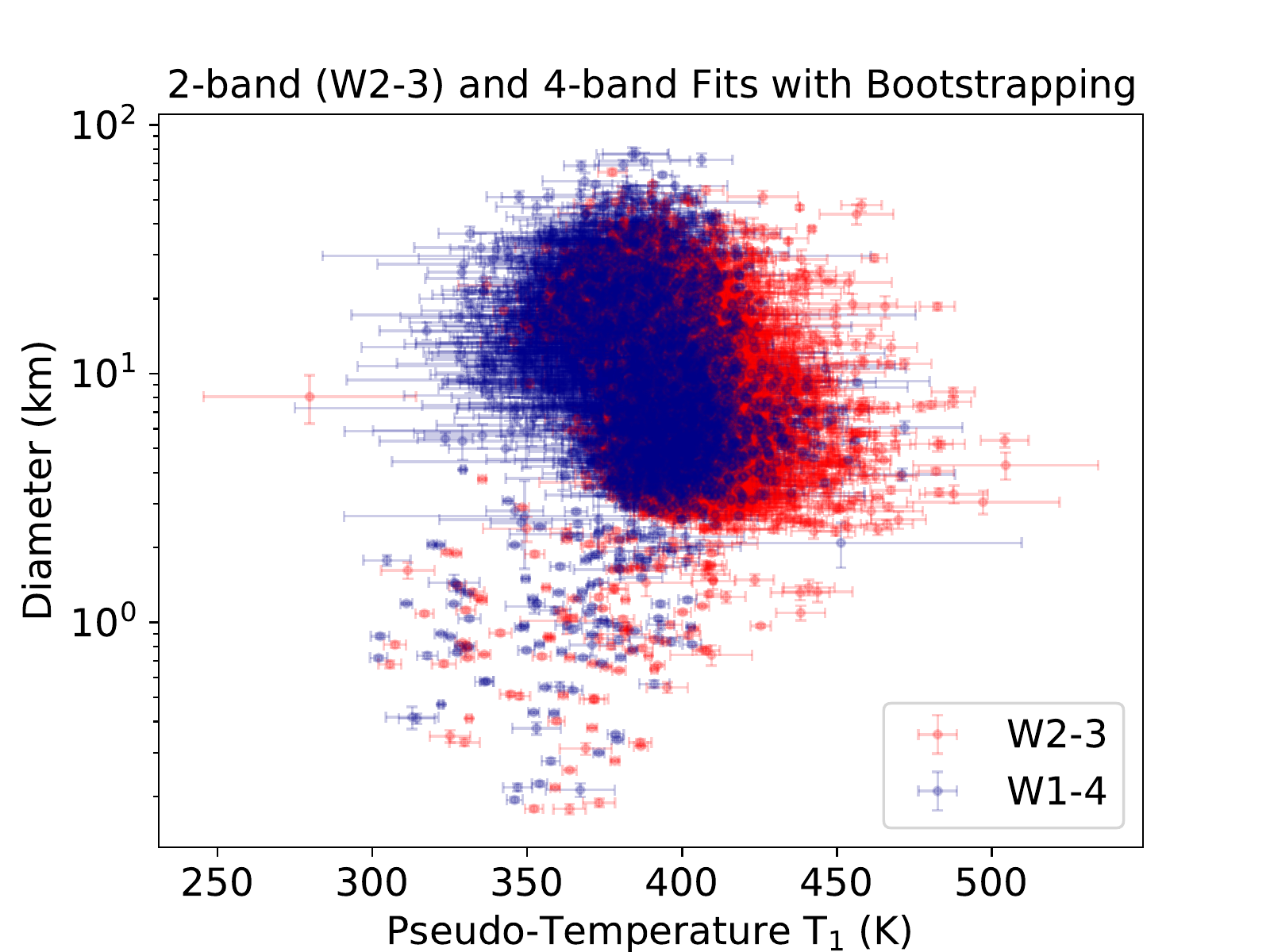} %
\caption{Diameter vs.\ pseudo-temperature $T_1$ for 4542 cases calculated from a two-band (W2--3) least-squares solution and four-band (W1--4) regularized solution to a reparameterized NEATM model.
  Each point represents the median diameter and pseudo-temperature value among all accepted bootstrap trials for an object while error bars represent the standard deviations of the diameter and pseudo-temperature values among accepted bootstrap trials.  The W2--3 data shown in this figure is available as the data behind the figure. The data also includes additional information that will allow the reader to recreate Figures 4--6.
}
\label{fig:TD_bootstrap}
\end{figure}

\begin{figure}
\centering\includegraphics[width=7cm]{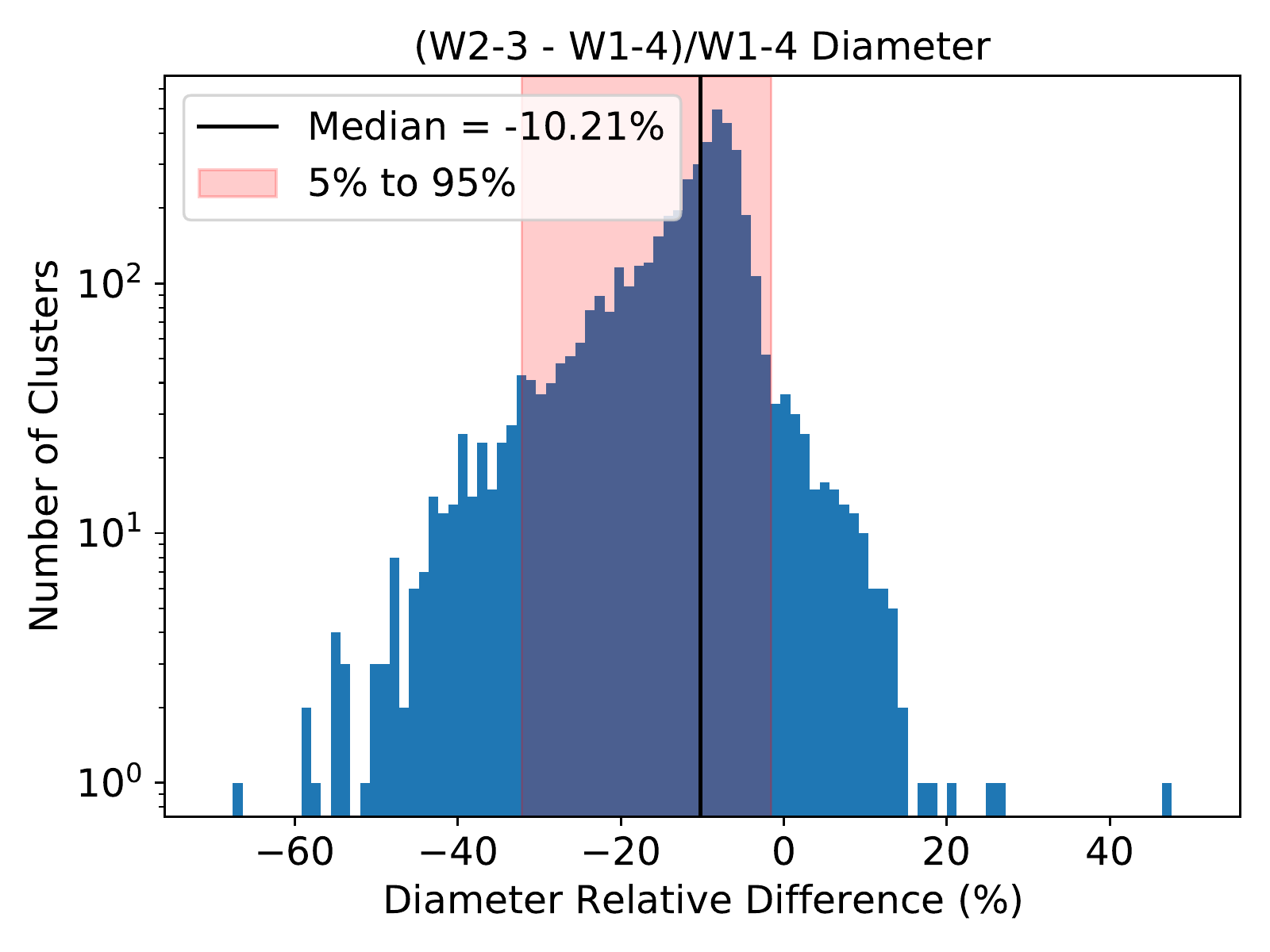} %
\centering\includegraphics[width=7cm]{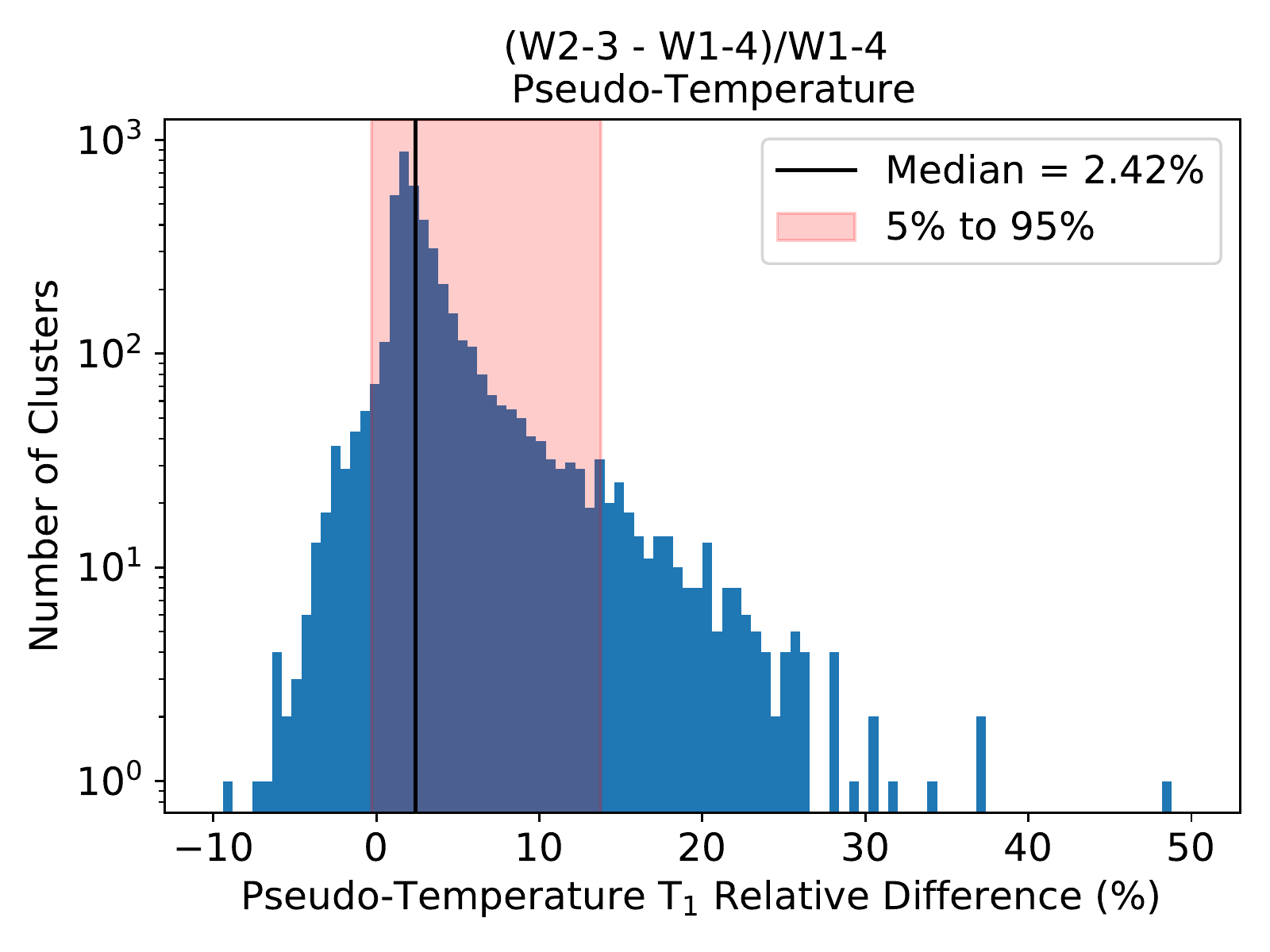}
\centering\includegraphics[width=7cm]{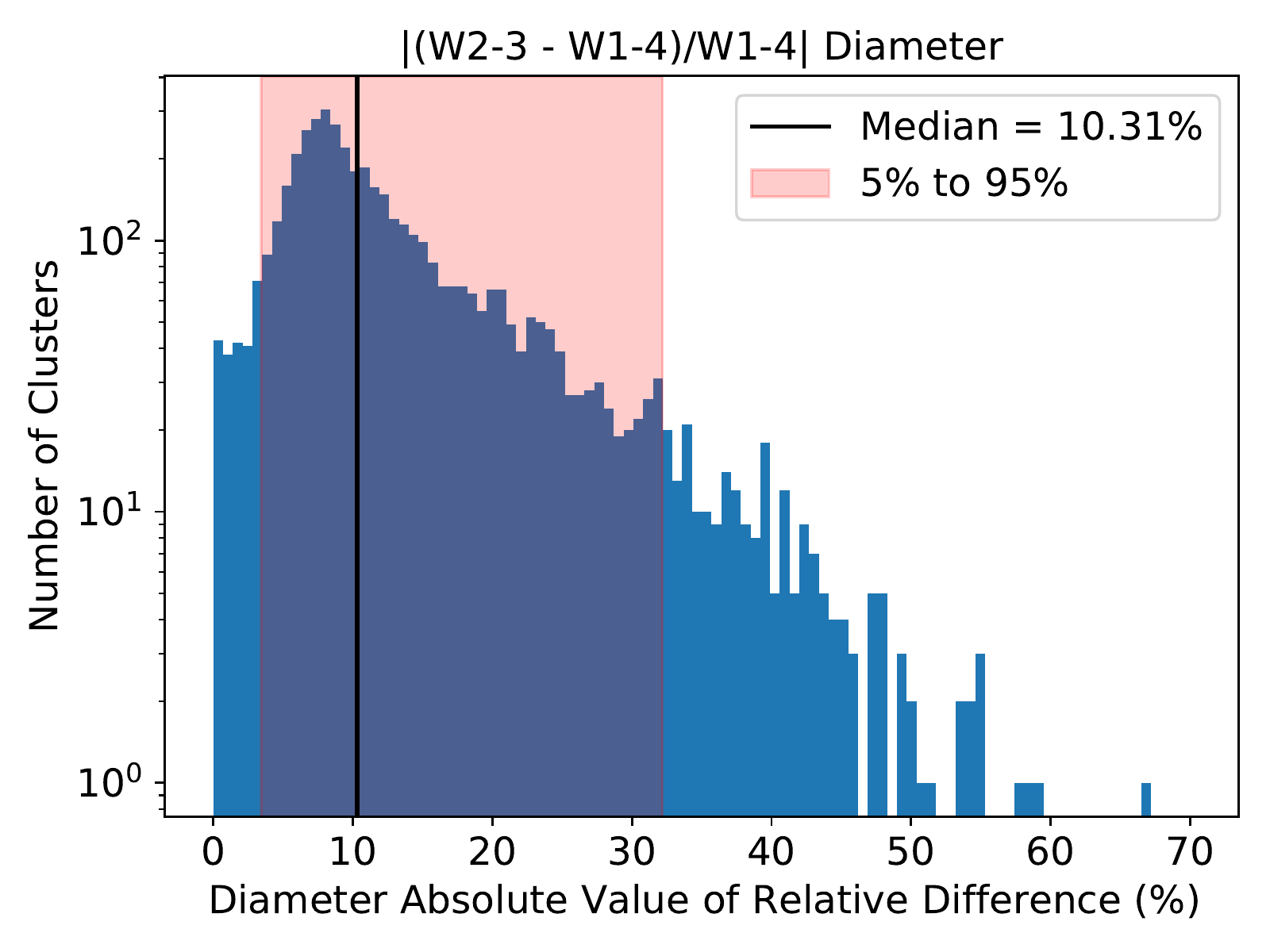}
\centering\includegraphics[width=7cm]{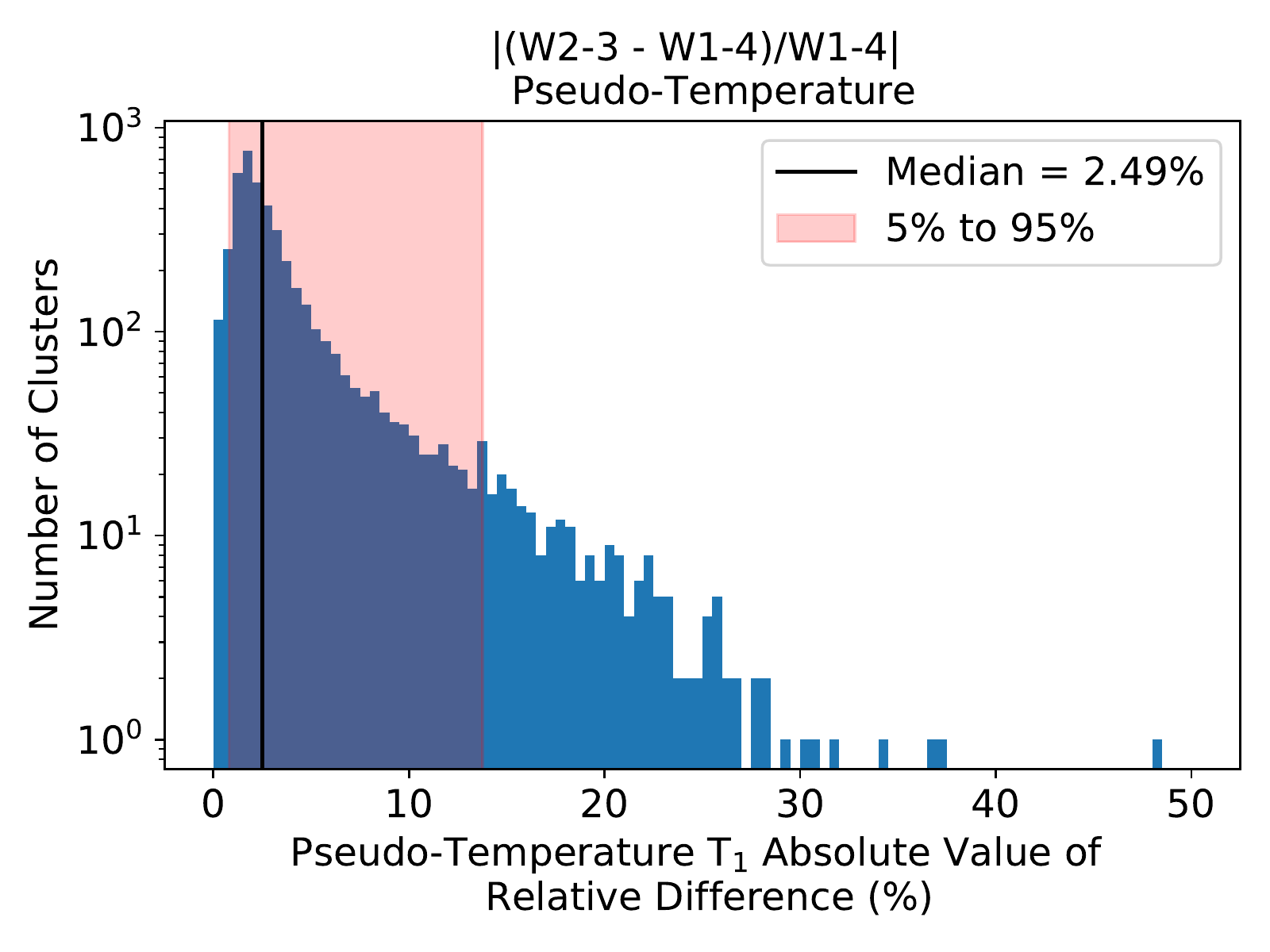}
\caption{Histograms of the relative difference and absolute value of the relative difference in $D$ and $T_1$ parameter estimates.  Median values are represented with a vertical black line while a 
pink highlighted block is used to represent the 5 -- 95\% quantile range.  Top Left: Relative difference in diameter. Top Right: Relative difference 
in pseudo-temperature $T_1$.  Bottom Left: Absolute value of relative difference in diameter. Bottom Right: Absolute value of relative difference 
in pseudo-temperature $T_1$.}
\label{fig:TD_reldiff}
\end{figure}

Figure \ref{fig:TD_bootstrap} shows the two-band (W2--3) and \citet{myhr22} four-band (W1--4) estimates of pseudo-temperature $T_1$ vs. diameter for 4542 clusters representing 4277 asteroids.  While the overlap between the two-band and four-band fits is substantial, there are also some differences worth noting.  For example, the two-band fits sway towards higher pseudo-temperatures and smaller diameters overall in comparison to their four-band counterparts (Figure \ref{fig:TD_reldiff}).
These percentage differences were obtained by calculating $(D_{\rm W2-3} - D_{\rm W1-4})/D_{\rm W1-4}$ for each cluster.
The median diameter difference is -10\%,
with the 5\% and 95\% quantiles of the distribution at -32\% and -1.5\%, respectively.  The median pseudo-temperature difference is 2.4\%, with 5\% and 95\% quantiles at -0.30\% and 14\% respectively).  
Of these results, 85 of the asteroids were near-Earth objects, representing 92 clusters.  Looking only at the near-Earth objects, the median diameter difference is 8.9\%,
with the 5\% and 95\% quantiles of the distribution at -20\% and -2.4\%, respectively.  The near-Earth object median pseudo-temperature difference is 2.1\%, with 5\% and 95\% quantiles at 0.57\% and 5.5\%, respectively.

\begin{figure}
\centering\includegraphics[width=7cm]{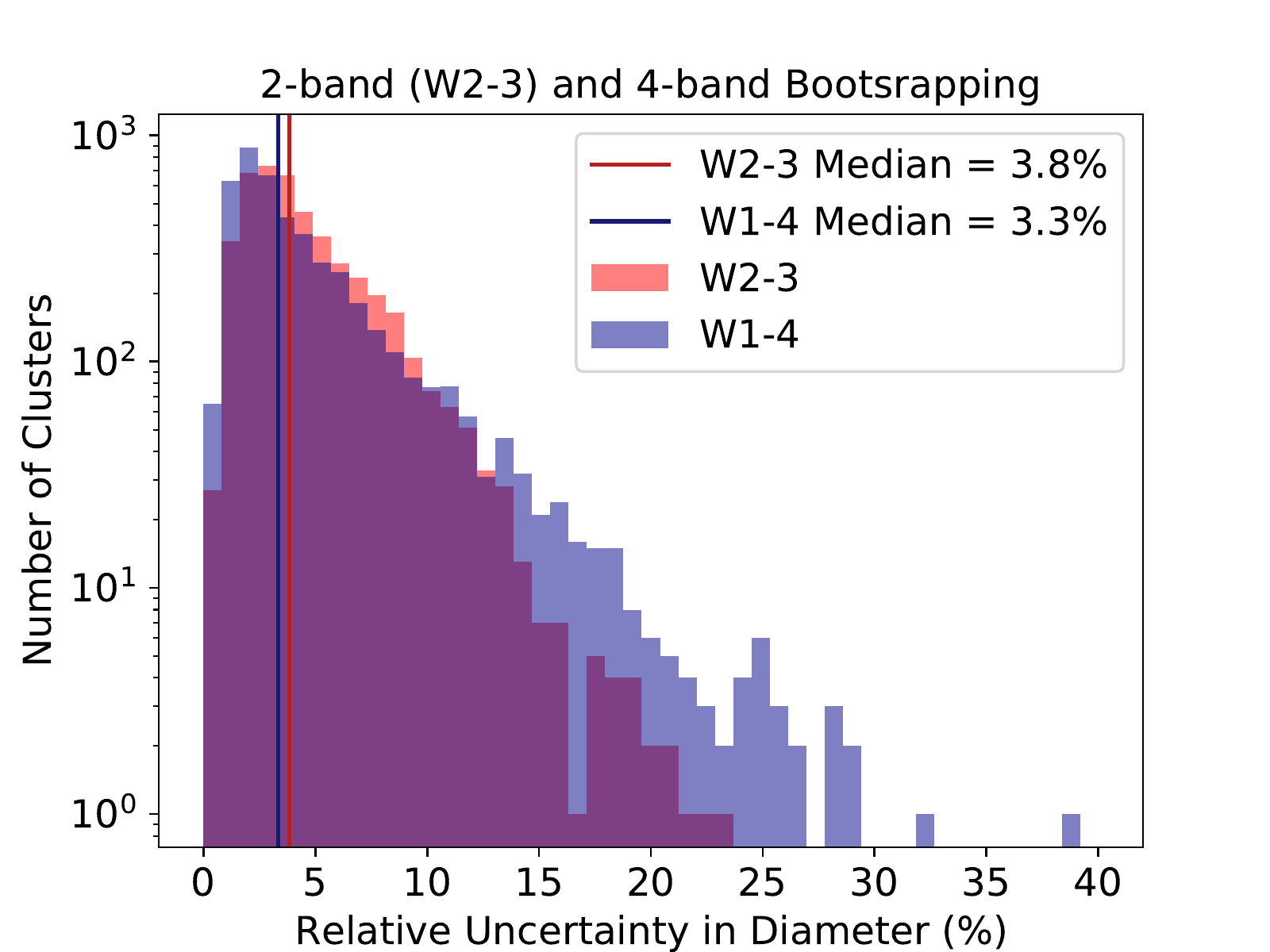}
  \centering\includegraphics[width=7cm]{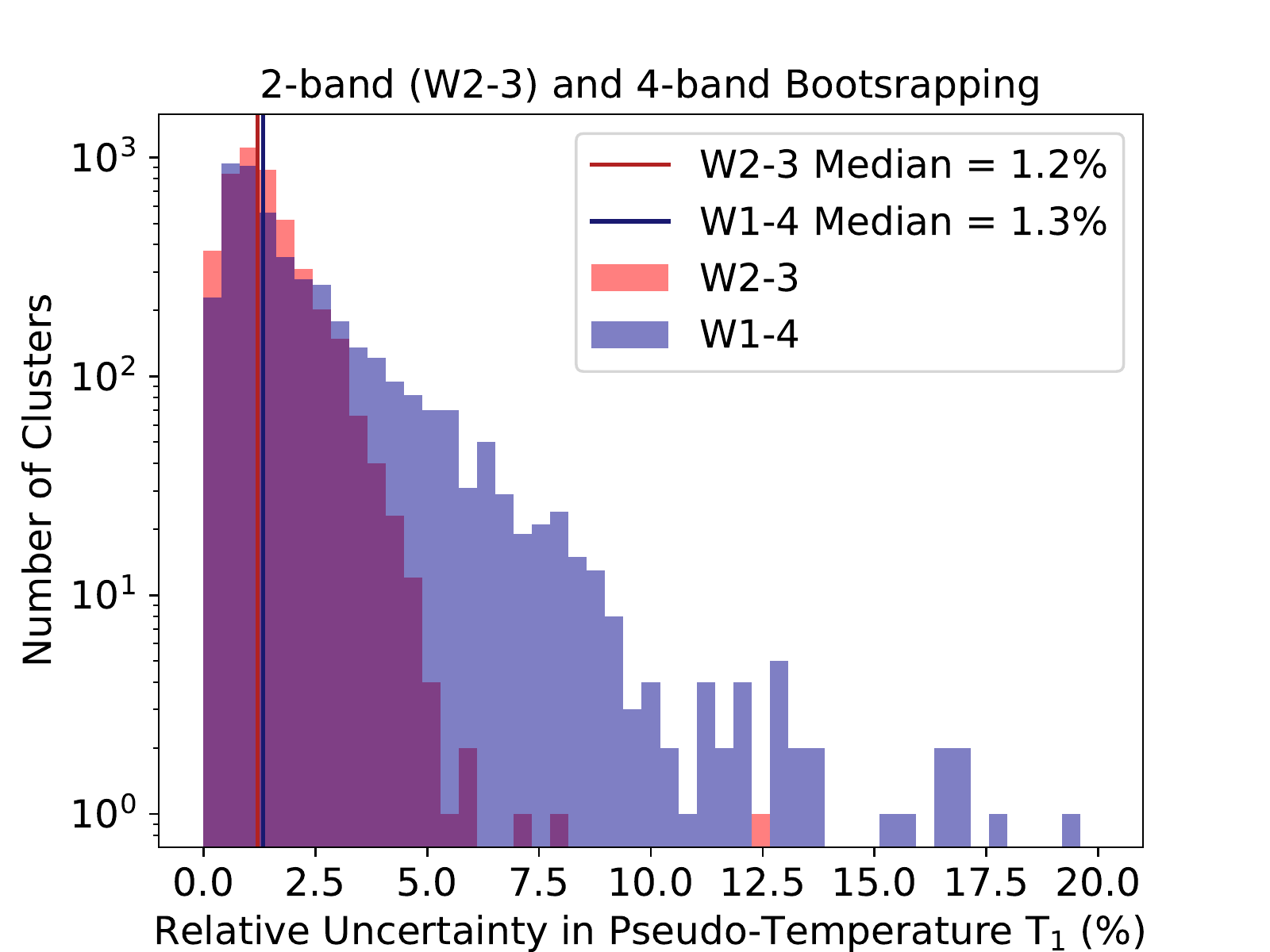}
\caption{Histograms of $D$ and $T_1$ parameter relative uncertainty, i.e., the standard deviation in accepted bootstrap trials for an object divided by the median value of the bootstrap trials.  The median of the relative uncertainties for all objects is shown as a vertical line.  Two-band results are shown in red whereas four-band results are 
shown in blue.  Left: Relative uncertainty in diameter. Right: Relative uncertainty in pseudo-temperature $T_1$.  }
\label{fig:TD_sigma}
\end{figure}

We found that two-band and four-band fits generally have comparable parameter uncertainties, although the median diameter uncertainty is somewhat larger for two-band fits (Figure \ref{fig:TD_sigma}).

\begin{figure}
\centering\includegraphics[width=7cm]{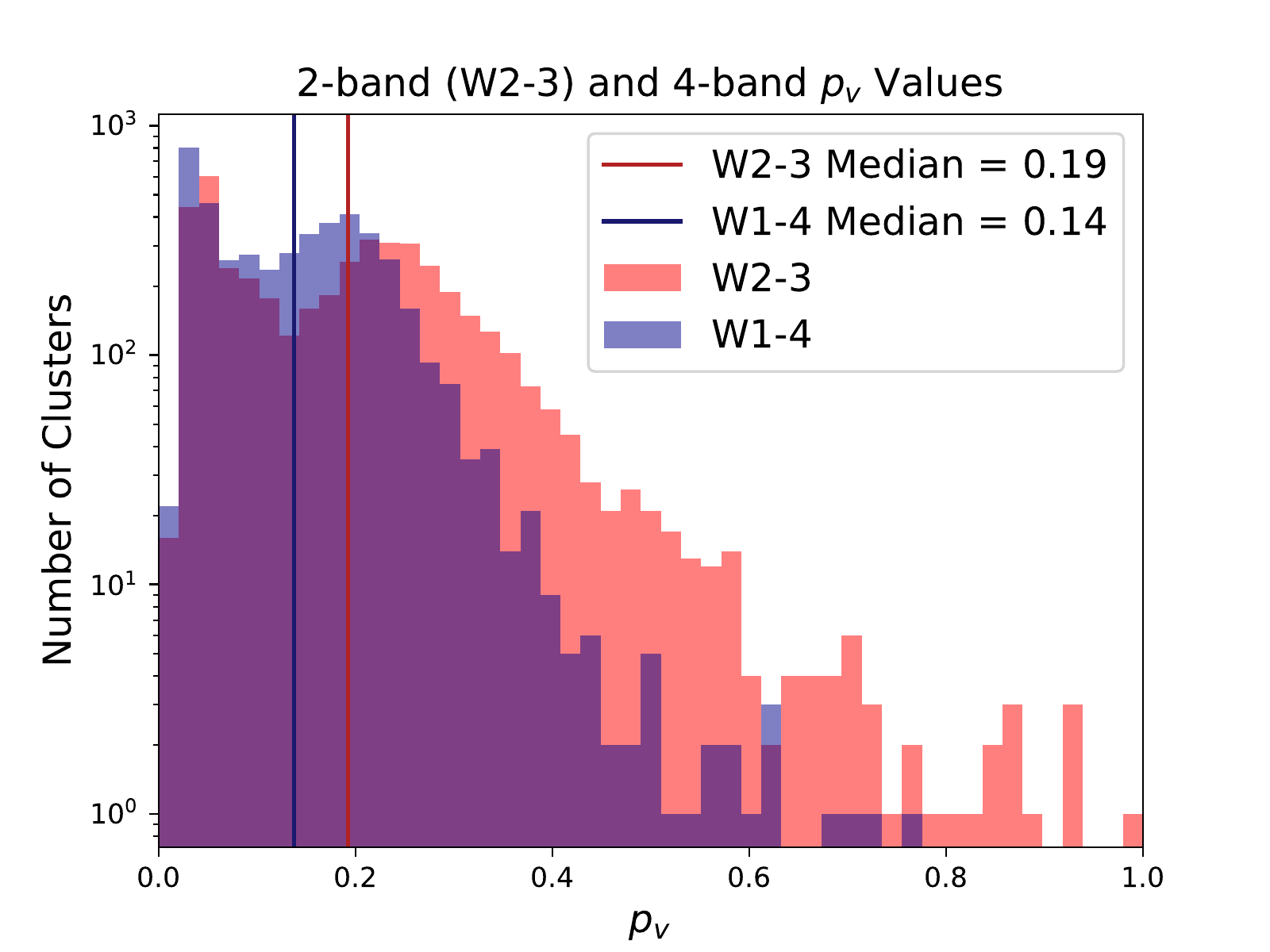}
\caption{Histogram of visible-band geometric albedos ($p_v$) as calculated using the median modeled diameter for each cluster and the absolute magnitude for the object, $H$.  The median albedo for the two-band fits is shown as a dark red line, while the median albedo of the four-band fits is shown as a dark blue line.}
\label{fig:albedos}
\end{figure}

Using equation \ref{eqn:albedo}, we also calculated the albedo for the two-band fits.  We eliminated any unphysical albedo values ($p_v >$ 1) from the dataset used to calculate the median albedo, resulting in 14 eliminated two-band clusters and 1 eliminated four-band cluster.  The results
(Figure \ref{fig:albedos}) demonstrate that the albedos for the W2--3 fits were 24\% larger than those of the W1--4 fits (median value, 95\% CI 2.9\% to 120\%).  The albedo uncertainty for the W2--3 fits was larger than for the W1--4 fits by 11.0\%, a surprisingly small increase explained by the uncertainty on H dominating the albedo error budget \citep{myhr22}.

\begin{figure}
\centering\includegraphics[height=7cm]{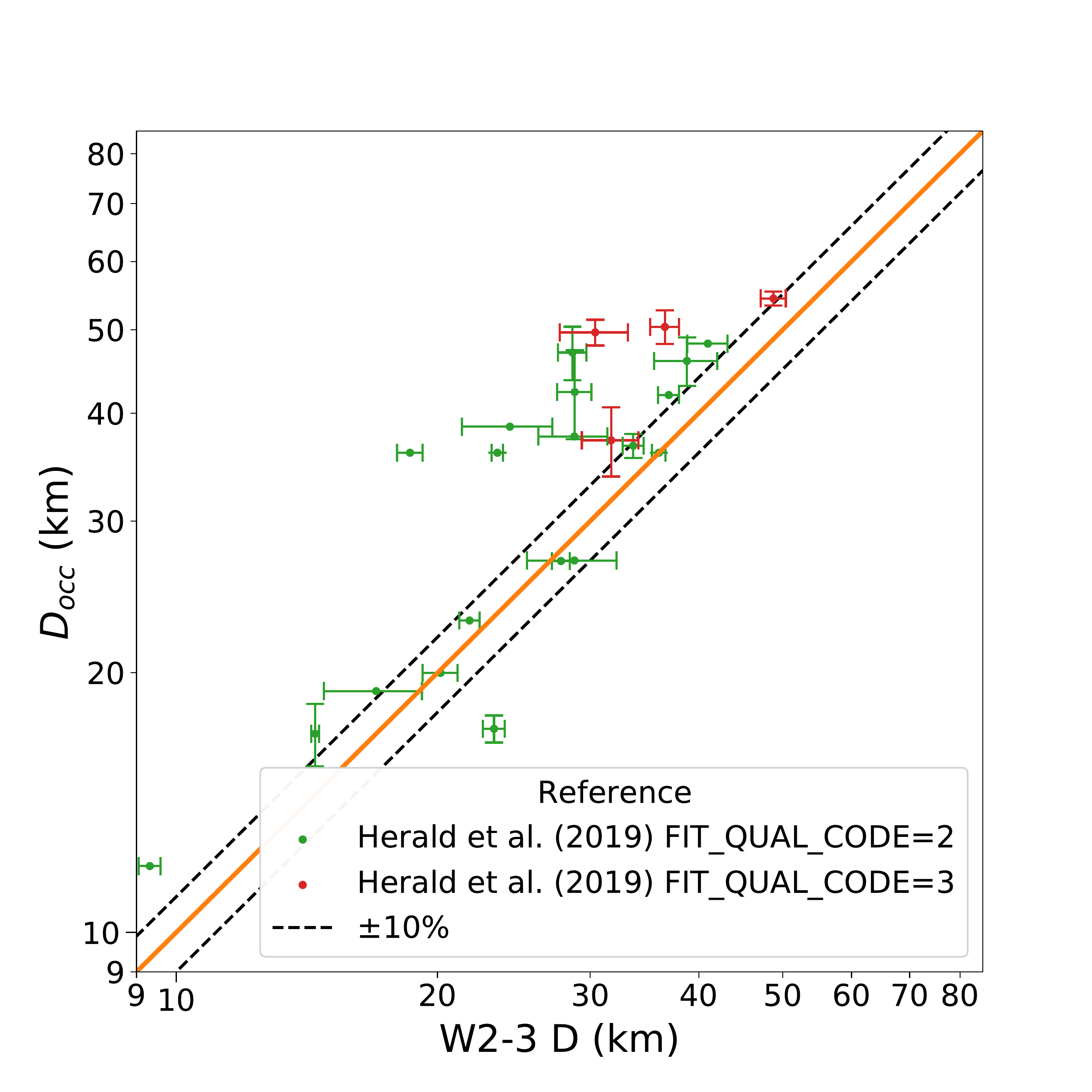}
\caption{Plot of occultation diameters versus our corresponding two-band (W2--3) modeled diameters.
  The orange line represents a perfect one-to-one ratio between diameters, whereas the black dotted lines represent a 10\% error in the positive and negative directions.}
\label{fig:occ}
\end{figure}

In addition to comparing two-band and four-band fit results, we were also interested in comparing two-band fit results to independent diameter estimates.  Short of visiting an asteroid with a spacecraft, the most accurate diameter measurements are obtained with radar observations or stellar occultation observations.  Here, we used the stellar occultation diameter estimates with quality codes of 2 or 3 \citep{herald19} to obtain 23 diameter comparisons for 22 distinct asteroids (Figure \ref{fig:occ}).  The two-band W2--3 results give generally smaller diameters than the occultation data and have a median error, obtained from the absolute value of the relative differences,
at 15\%, with a worst case error of 39\%.  This comparison further confirms our finding that two-band diameters are $\sim$10\% smaller than their true diameters.  

\begin{figure}
\centering\includegraphics[height=7cm]{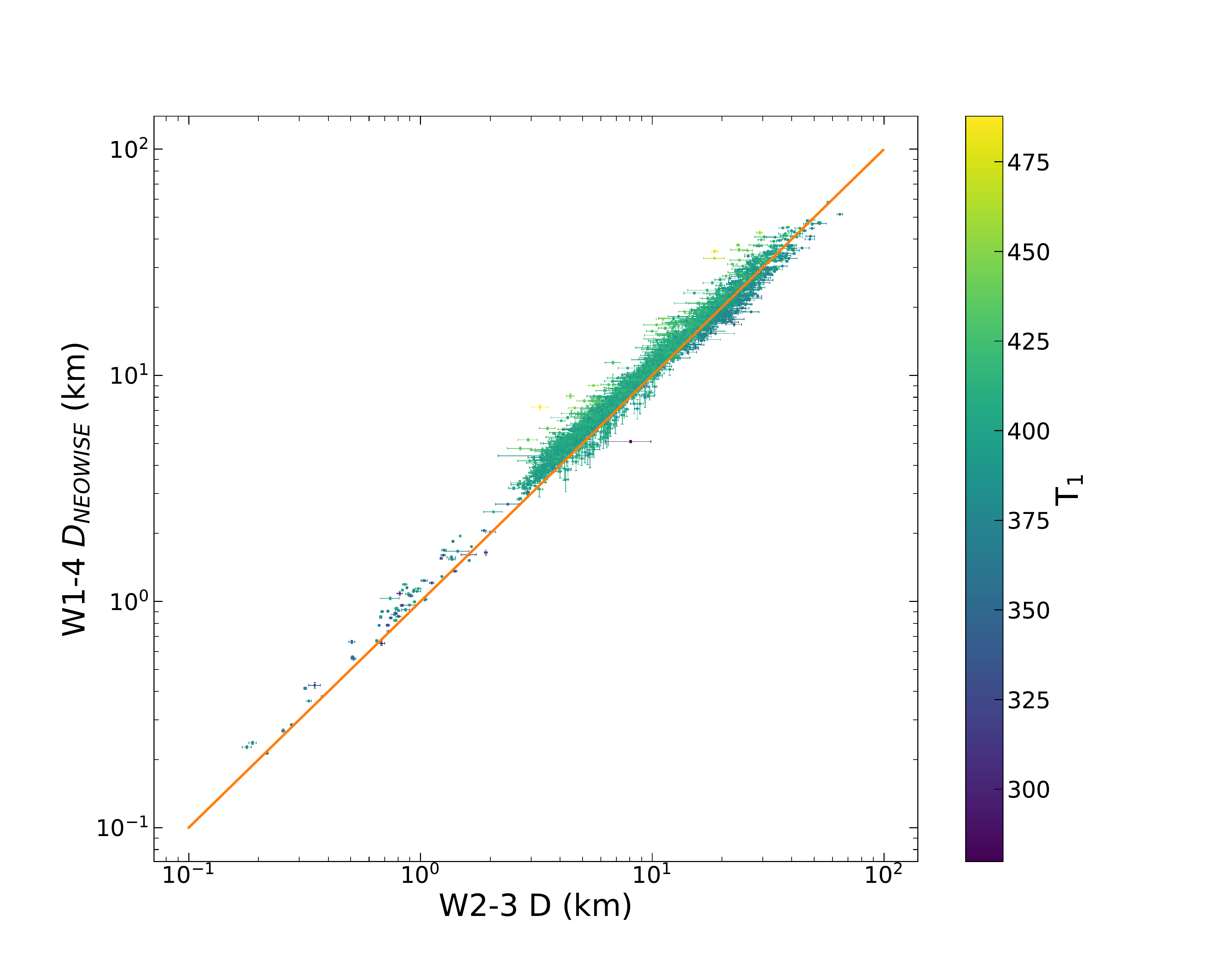}
\centering\includegraphics[height=7cm]{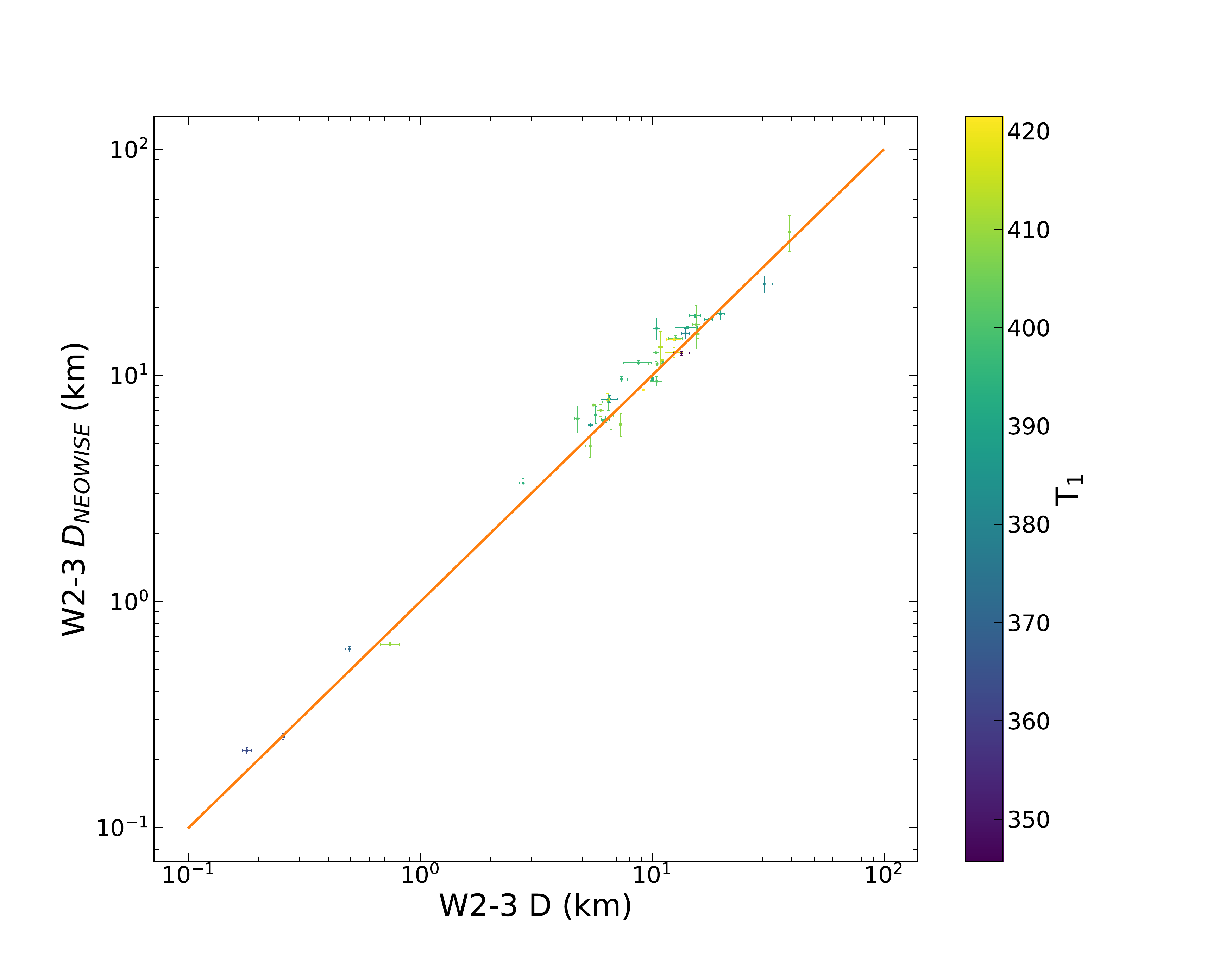}
\caption{Left: Plot of NEOWISE four-band diameters versus our corresponding two-band (W2--3) modeled diameters.  The orange line represents a perfect one-to-one ratio between diameters.  
  The color bar indicates our pseudo-temperature modeled with W2 and W3.  
  Right: Plot of NEOWISE two-band (W2--3) diameters versus our corresponding two-band (W2--3) modeled diameters.}
\label{fig:neowise}
\end{figure}

Finally, we also compared our two-band diameter estimates to two-band and four-band diameter estimates from NEOWISE
\citep{main19dps}.
For the four-band results, we found 2632 asteroids in common, where 137 asteroids were represented in two different epochs, which therefore resulted in 2769 diameter comparisons (Figure \ref{fig:neowise}, left).
The median percent difference
$(D_{\rm W2-3} - D_{\rm W1-4,N})/D_{\rm W1-4,N}$
was -7.8\% (5--95\% quantiles at -22\% and 11\%), suggesting that the NEOWISE four-band diameter estimates are also underestimates of the true diameter, but to a lesser degree than our two-band fits.
This trend remained true when we compared our two-band results to the NEOWISE two-band (W2--3) results.  With 42 asteroids in common and 0 asteroids represented in two epoch clusters, giving 42 diameter comparisons, the median percent difference
$(D_{\rm W2-3} - D_{\rm W2-3,N})/D_{\rm W2-3,N}$
was -9.1\% (5--95\% quantiles at -25\% and 15\%).

\pagebreak
\subsection{W3--4 Results}

\begin{figure} 
\centering\includegraphics[width=.6\linewidth]{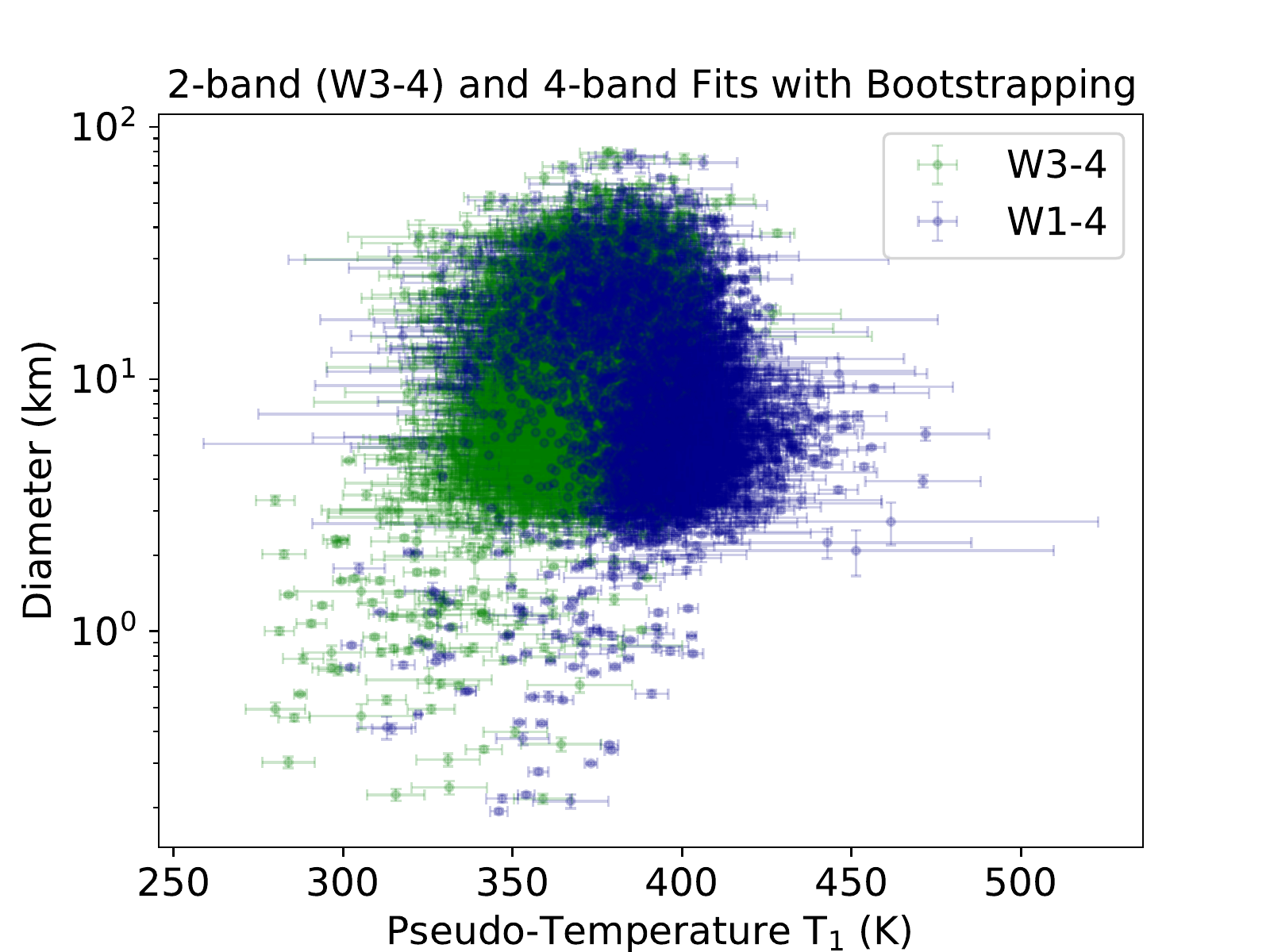}
\caption{Diameter vs.\ pseudo-temperature $T_1$ for 4685 cases calculated from a two-band (W3--4) and four-band (W1--4) reparameterized NEATM model.
  Each point represents the median diameter and pseudo-temperature value among all accepted bootstrap trials for an object while error bars represent the standard deviations of the diameter and pseudo-temperature values among accepted bootstrap trials.  The W3--4 data shown in this figure is available as the data behind the figure. The data also includes additional information that will allow the reader to recreate Figures 10--12.
}
\label{fig:TD_bootstrap_w3w4}
\end{figure}

\begin{figure}
\centering\includegraphics[width=7cm]{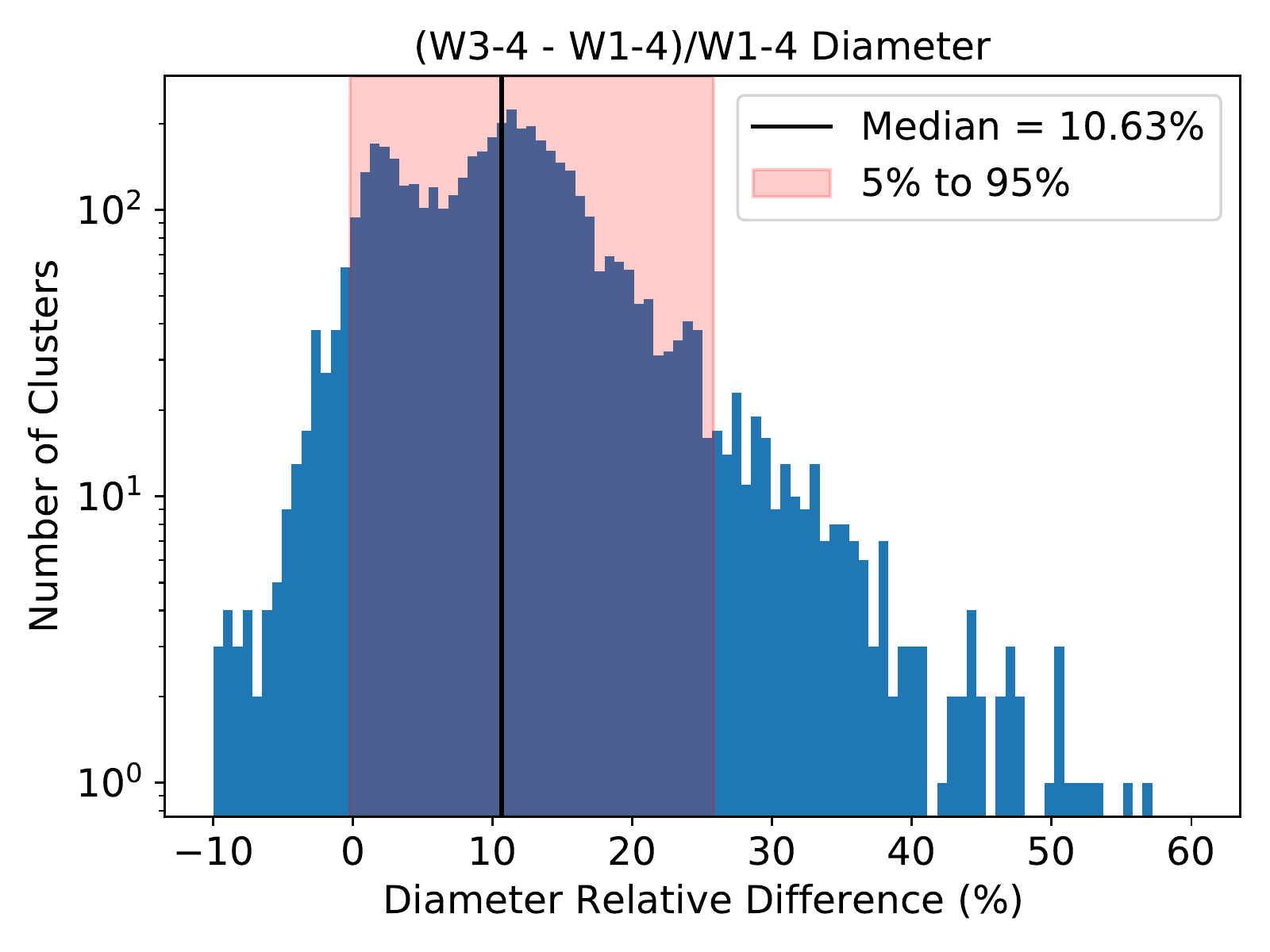}
\centering\includegraphics[width=7cm]{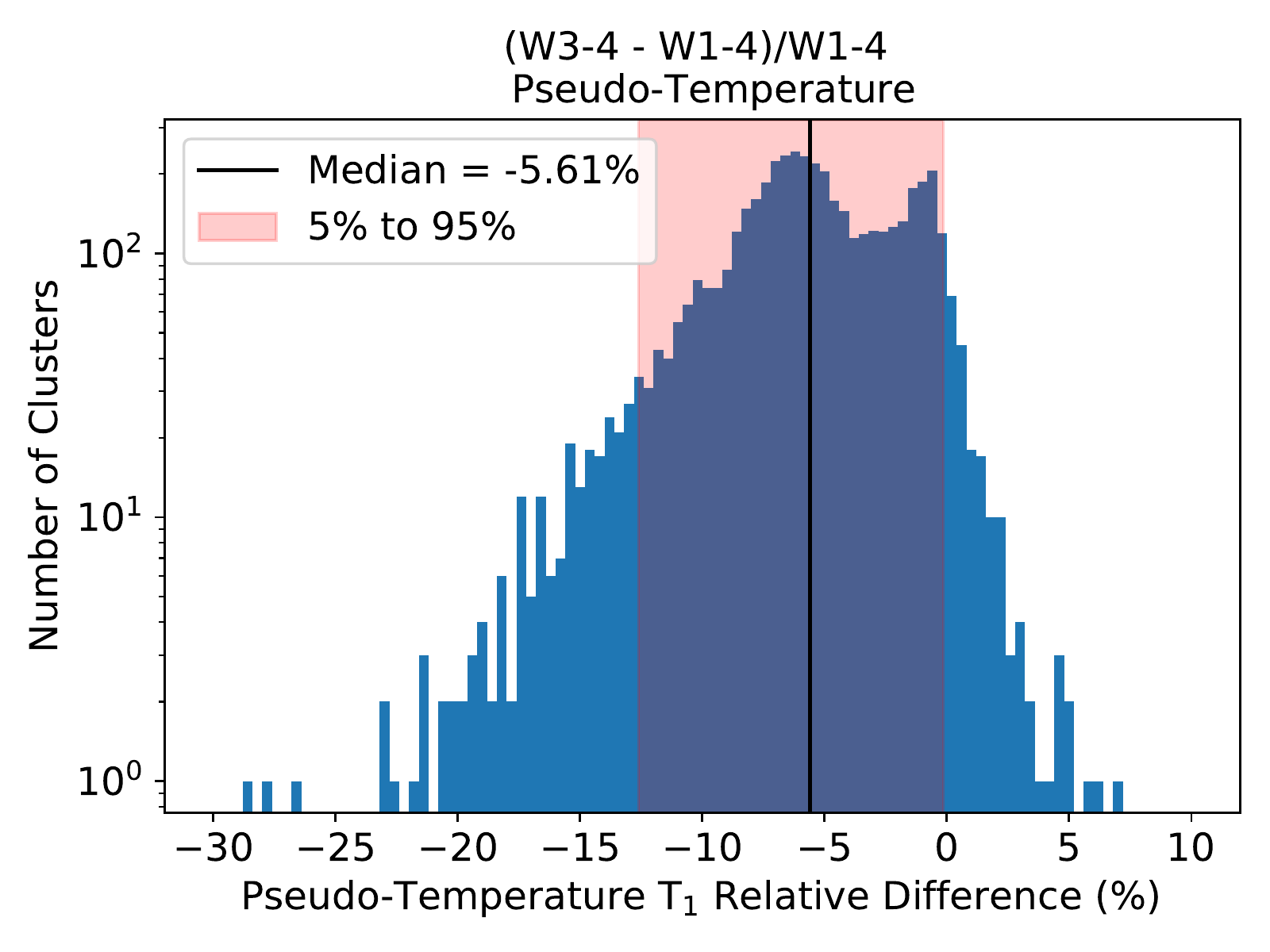}
\centering\includegraphics[width=7cm]{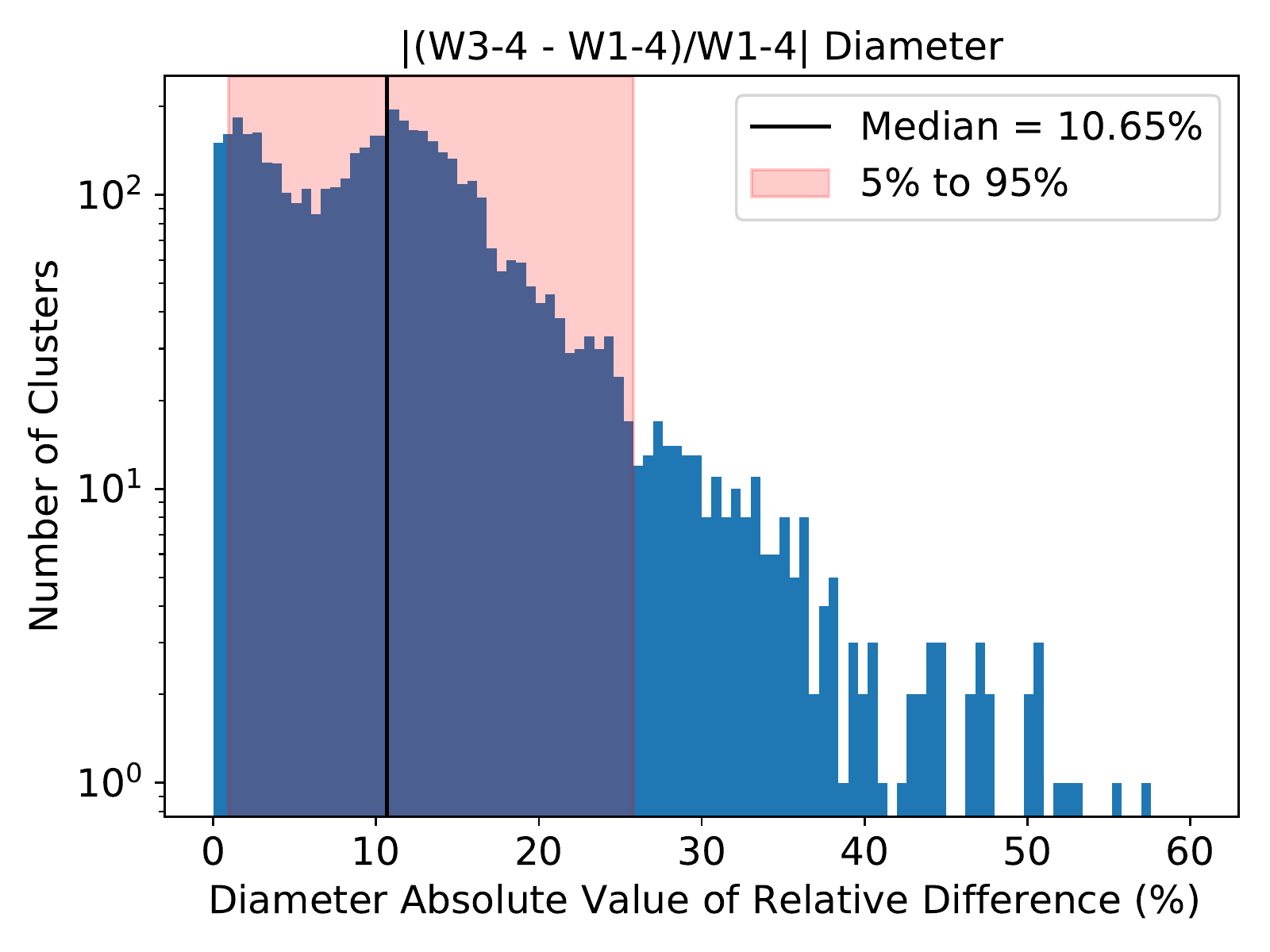}
\centering\includegraphics[width=7cm]{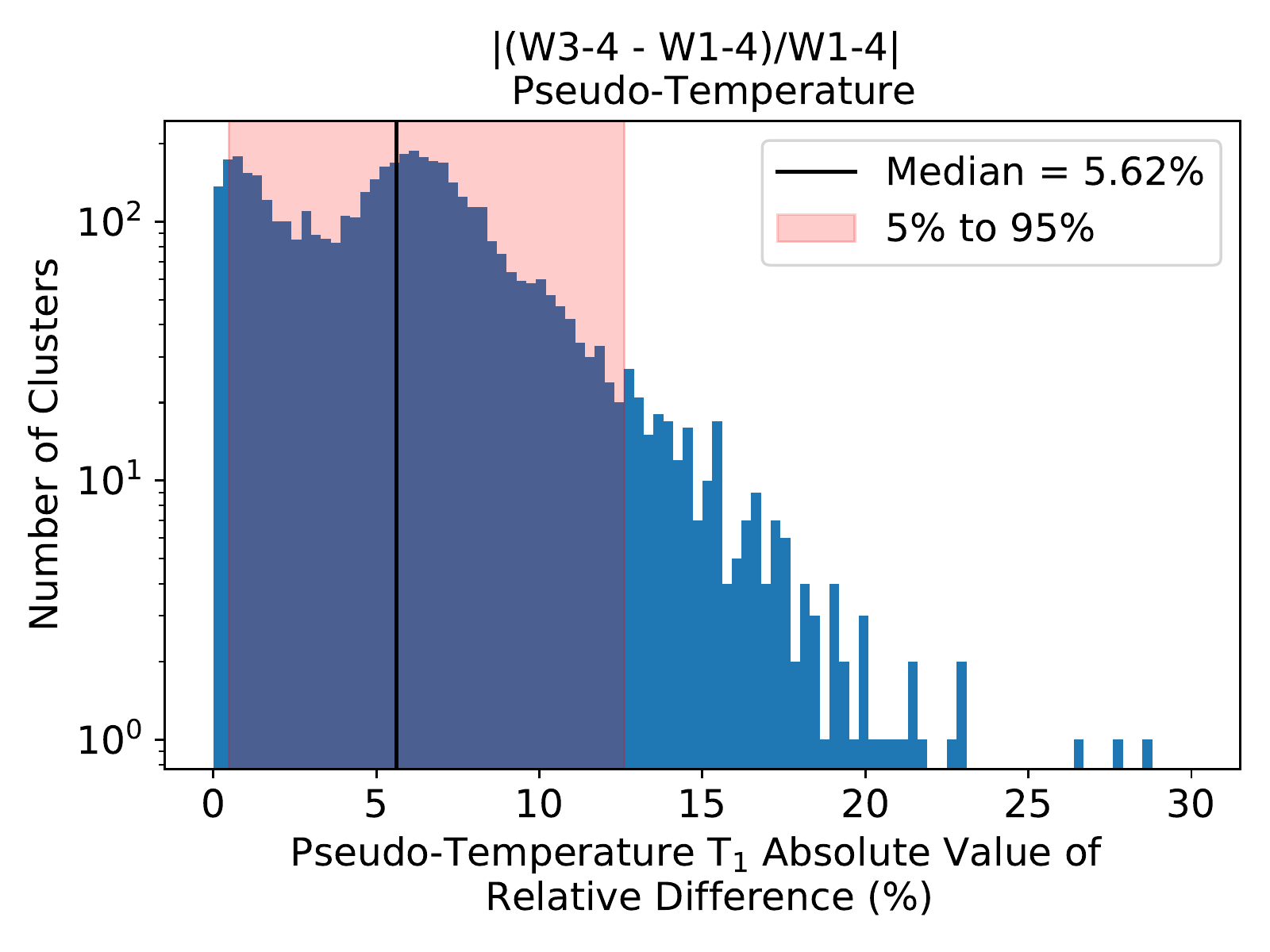}
\caption{Histograms of the relative difference and absolute value of the relative difference.  Median values are represented with a vertical black line while a 
pink highlighted block is used to represent the 5 - 95\% quantile range.  Top Left: Relative difference in diameter. Top Right: Relative difference 
in pseudo-temperature.  Bottom Left: Absolute value of relative difference in diameter. Bottom Right: Absolute value of relative difference 
in pseudo-temperature.}
\label{fig:TD_reldiff_w3w4}
\end{figure}

Figure \ref{fig:TD_bootstrap_w3w4} shows the two-band (W3--4) and \citet{myhr22} four-band (W1--4) estimates of pseudo-temperature $T_1$ vs.\ diameter for 4685 clusters, representing 4420 asteroids.  The overlap between the two-band (W2--3) and four-band fits is substantial.  However, the W3--4 fits appear to deviate from the four-band fits in the opposite directions as that of W2--3.  For example, the two-band (W3--4) fits tend towards smaller pseudo-temperatures and larger diameters overall in comparison to their four-band counterparts (Figure \ref{fig:TD_reldiff_w3w4}),
in contrast to the trends observed in W2--3.
The median diameter difference is 11\%, with
the 5\% and 95\% quantiles of the distribution at -2.1\% and 26\%, respectively.
The median pseudo-temperature difference is -5.6\%,
with the 5\% and 95\% quantiles of the distribution at -13\% and -0.15\%, respectively.
Of these results, 87 of the asteroids were near-Earth objects, representing 95 clusters.  Looking only at the near-Earth objects, the median diameter difference is 14\%, with the 5\% and 95\% quantiles of the distribution at 3.9\% and 24\%, respectively.  The median pseudo-temperature difference is -8.3\%,
with the 5\% and 95\% quantiles of the distribution at -13\% and -3.5\%, respectively. 

\begin{figure}
\centering\includegraphics[width=7cm]{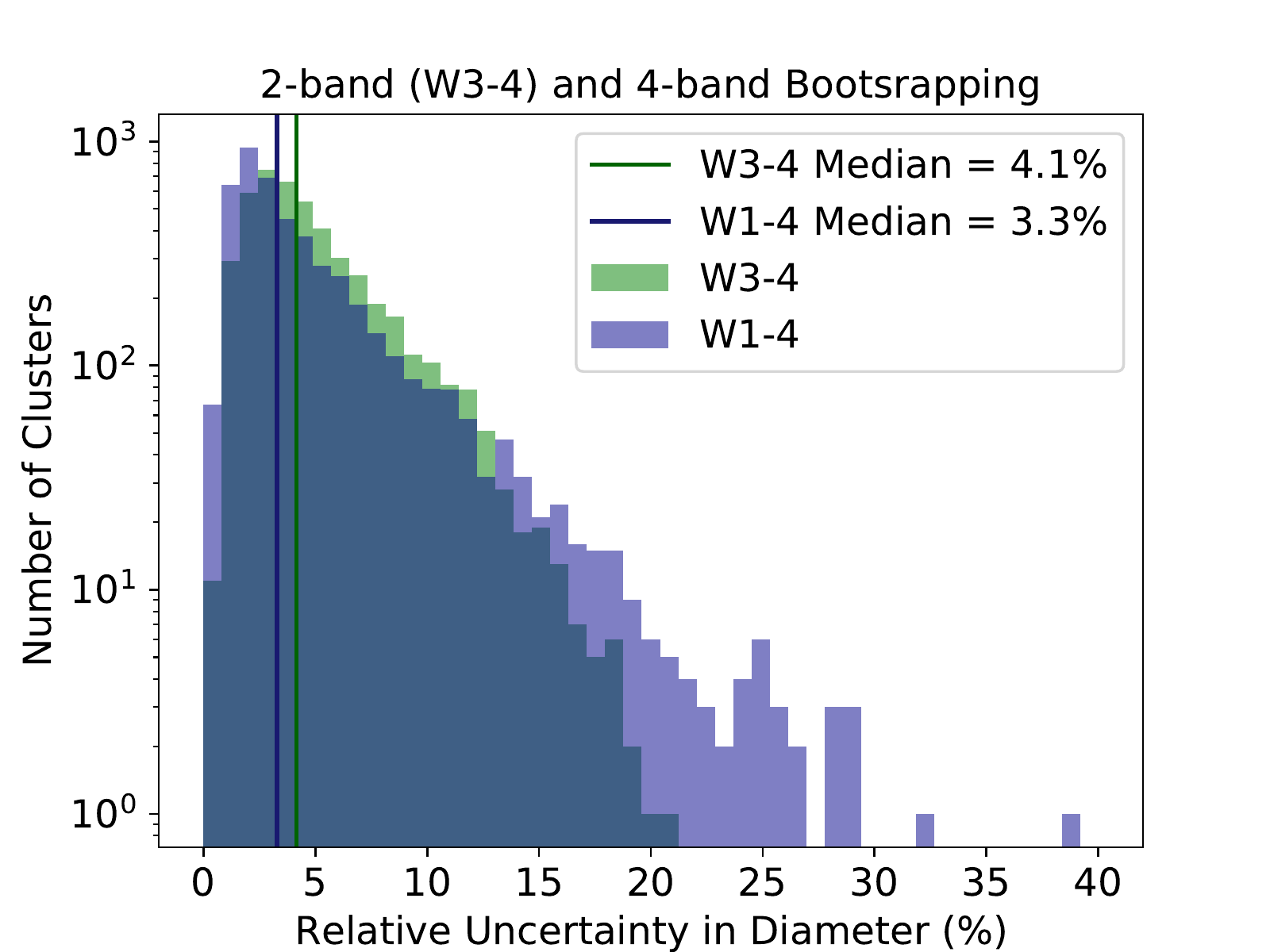}
\centering\includegraphics[width=7cm]{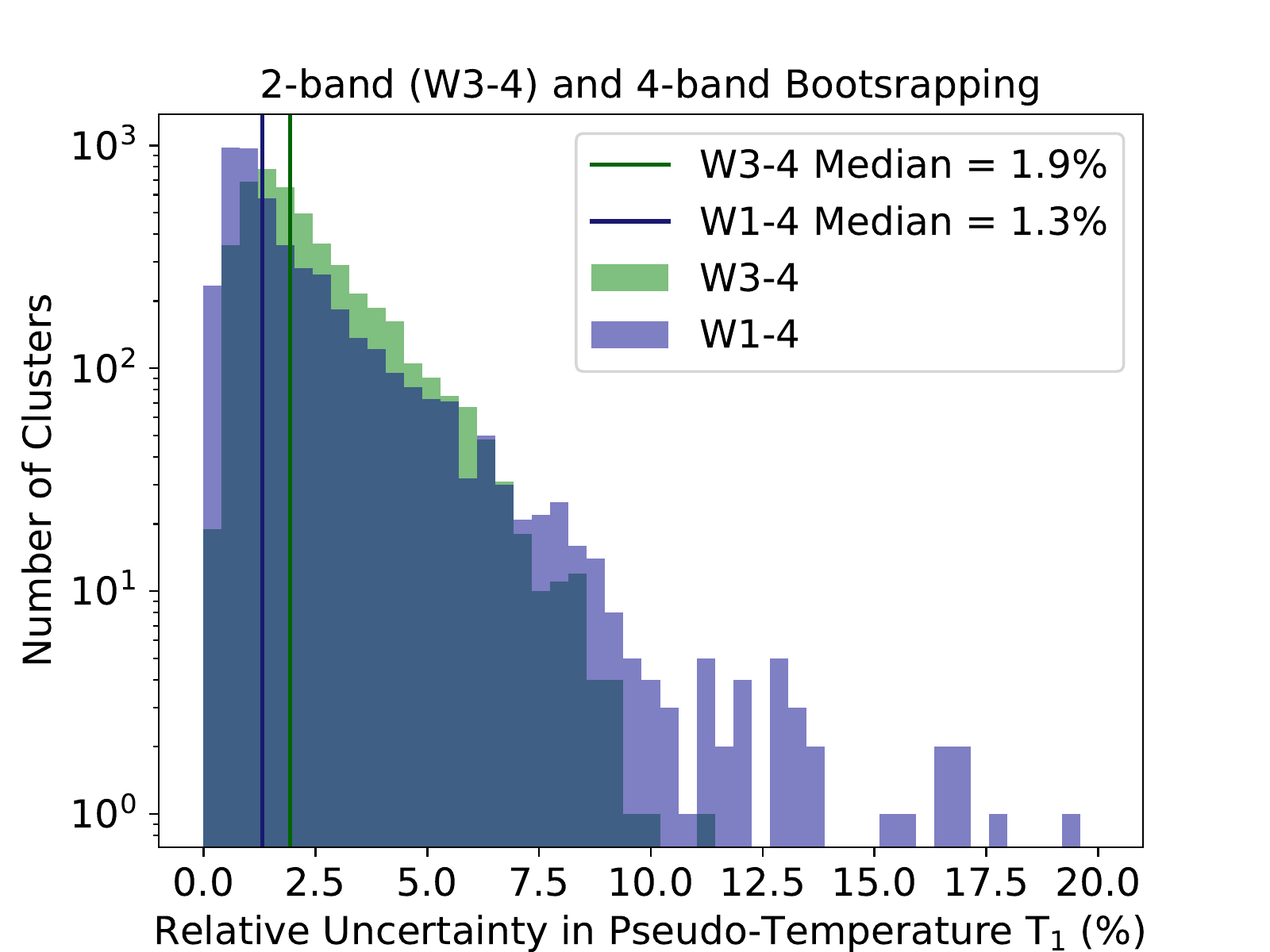}
\caption{Histograms of relative uncertainty, where relative uncertainty represents the standard deviation in accepted bootstrap trials for an object divided by the median value
of the bootstrap trials.  The median of the relative uncertainty for all objects is shown as a vertical line.  Two-band results are shown in green while four-band results are 
shown in blue.  Left: Relative uncertainty in diameter.  Right: Relative uncertainty in pseudo-temperature.}
\label{fig:TD_sigma_w3w4}
\end{figure}

We found that two-band fits have a slightly larger spread in parameter uncertainties compared to four-band fits (Figure \ref{fig:TD_sigma_w3w4}).

\begin{figure}
\centering\includegraphics[width=7cm]{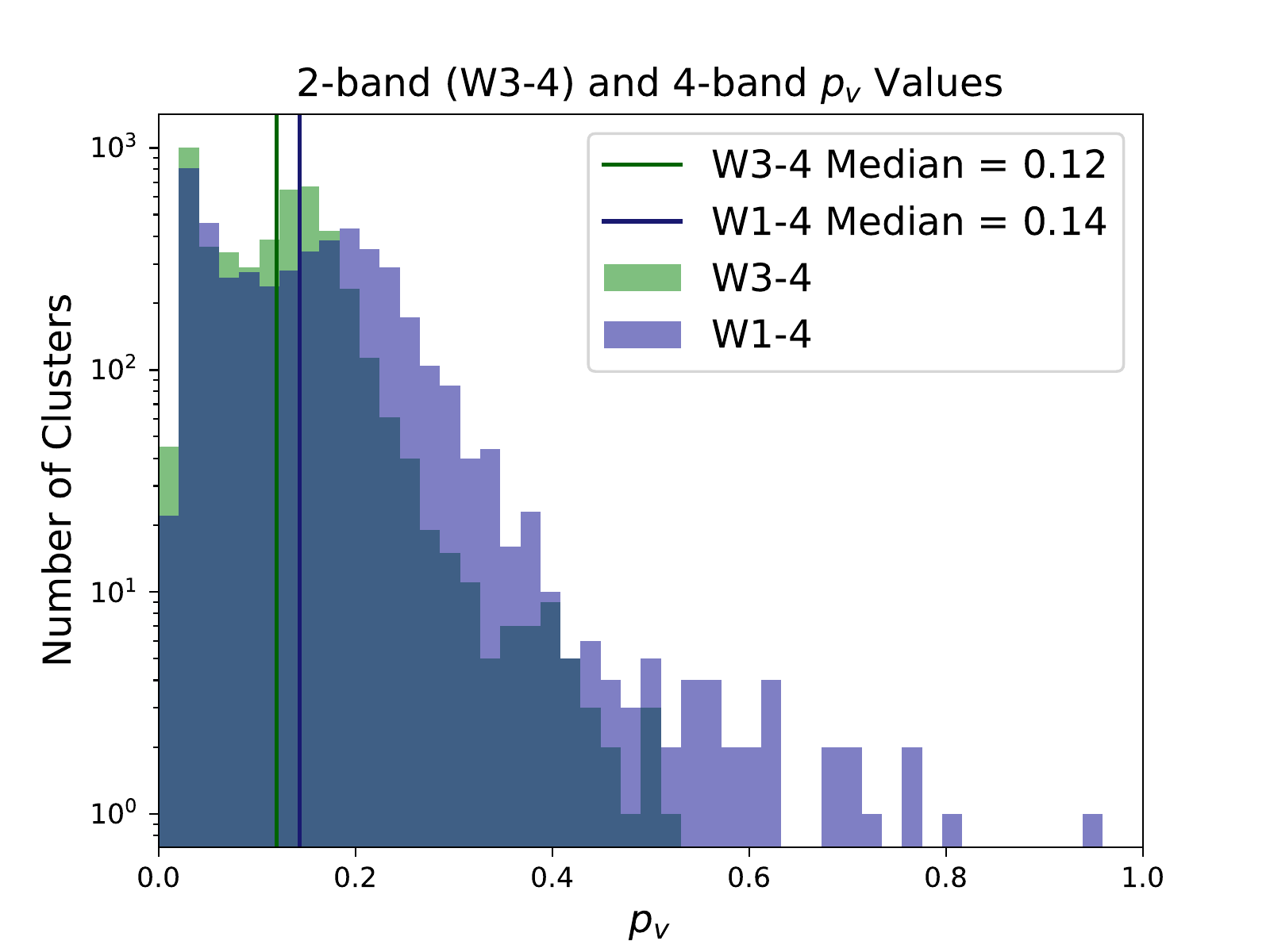}
\caption{Histogram of visible-band geometric albedos ($p_v$) as calculated using the median modeled diameter for each cluster and the absolute magnitude for the object, $H$.  The median albedo for the two-band fits is shown as a dark green line, while the median albedo of the four-band fits is shown as a dark blue line.}
\label{fig:albedos_w3w4}
\end{figure}

When obtaining the albedos for the W3--4 fits, there were no unphysical values which needed to be eliminated.  However, one unphysical albedo ($p_v >$ 1) was eliminated from the four-band dataset before calculating the median albedo.  The albedo distribution for W3--4 fits (Figure \ref{fig:albedos_w3w4}) indicate that the albedos for the two-band (W3--4) fits were 18\% smaller than those of the four-band (W1--4) fits (median value, 95\% CI -37\% to 42\%).  The albedo uncertainty for the W3--4 fits was larger than for the W1--4 fits by 1.2\%.

\begin{figure}
\centering\includegraphics[height=7cm]{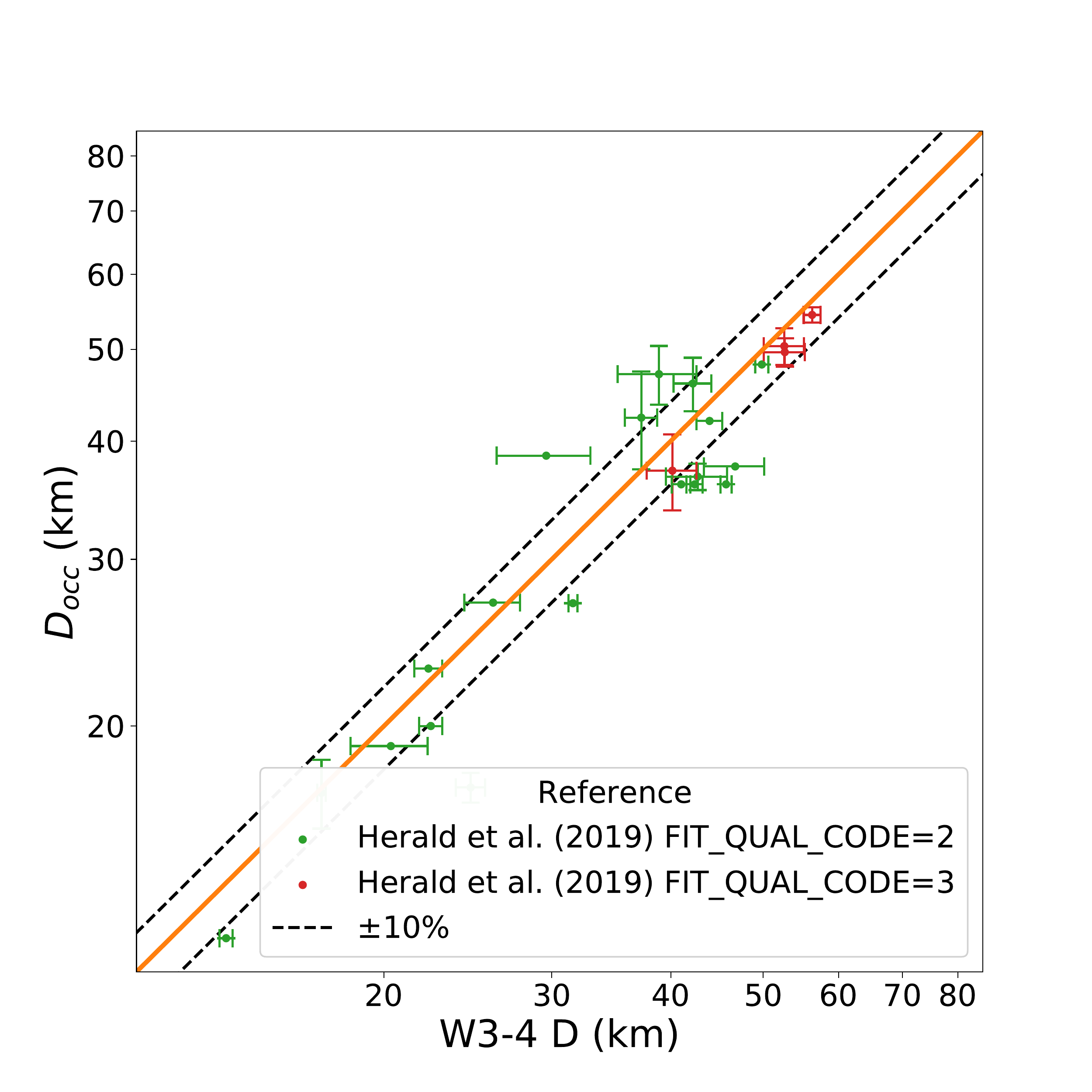}
\caption{Left: Plot of occultation diameters versus our corresponding diameters modeled with W3 and W4.  Red points represent objects with a higher quality occultation diameter with a quality code of 3, while green objects have a slightly lower quality occultation diameter with a quality code of 2.  The orange line represents a perfect one-to-one ratio between diameters, while black dotted lines represent a 10\% error in the positive and negative directions.}
\label{fig:occ_w3w4}
\end{figure}

We compared our diameter estimates to stellar occultation diameter estimates with quality codes of 2 or 3 \citep{herald19}.  With 23 diameter comparisons for 22 distinct asteroids (Figure \ref{fig:occ_w3w4}), we found that the W3--4 results usually give larger diameters than the occultation data and have a median error, obtained from the absolute value of the relative differences, at 12\%, with a worst case error of 27\%.

\begin{figure}
\centering\includegraphics[height=7cm]{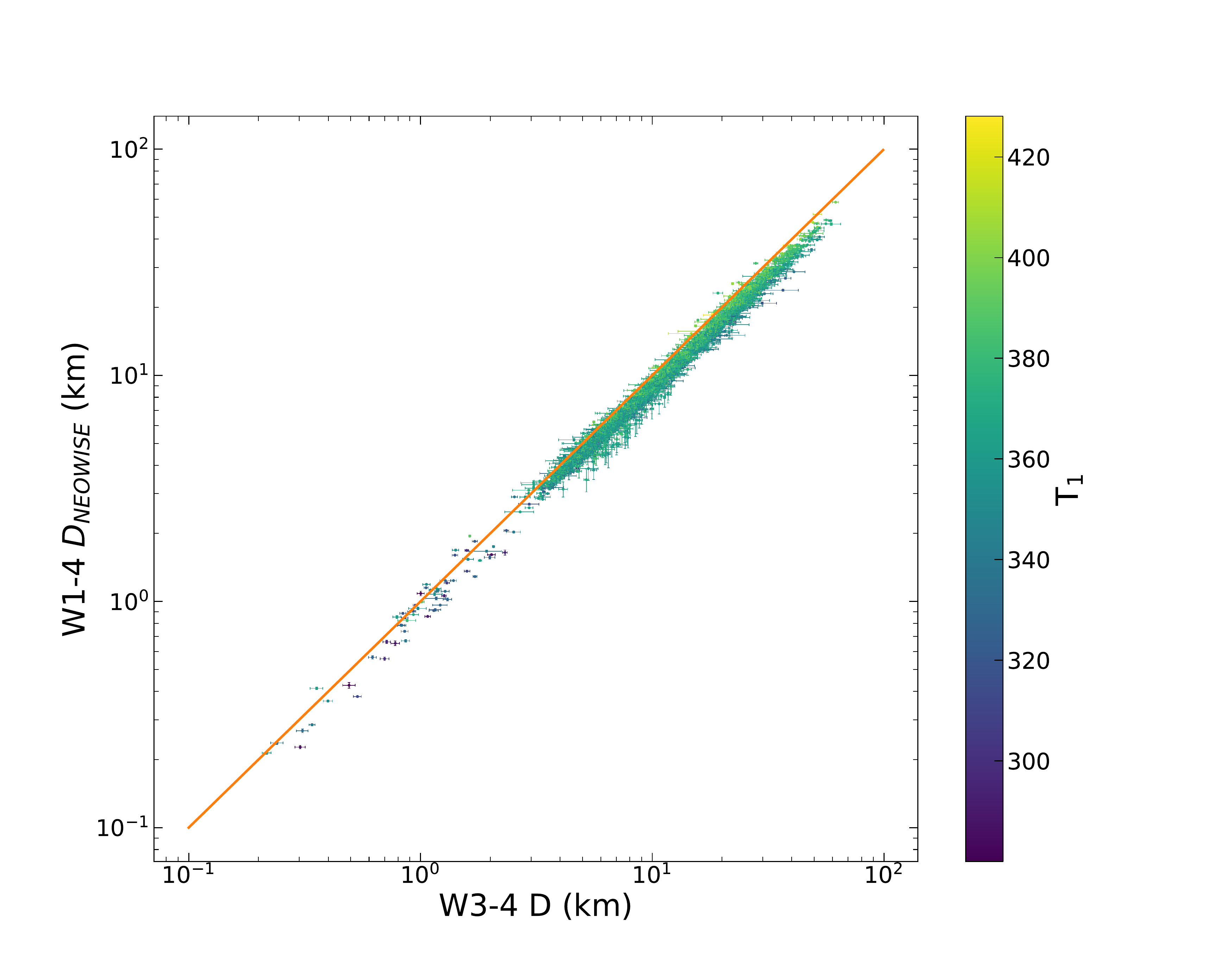}
\centering\includegraphics[height=7cm]{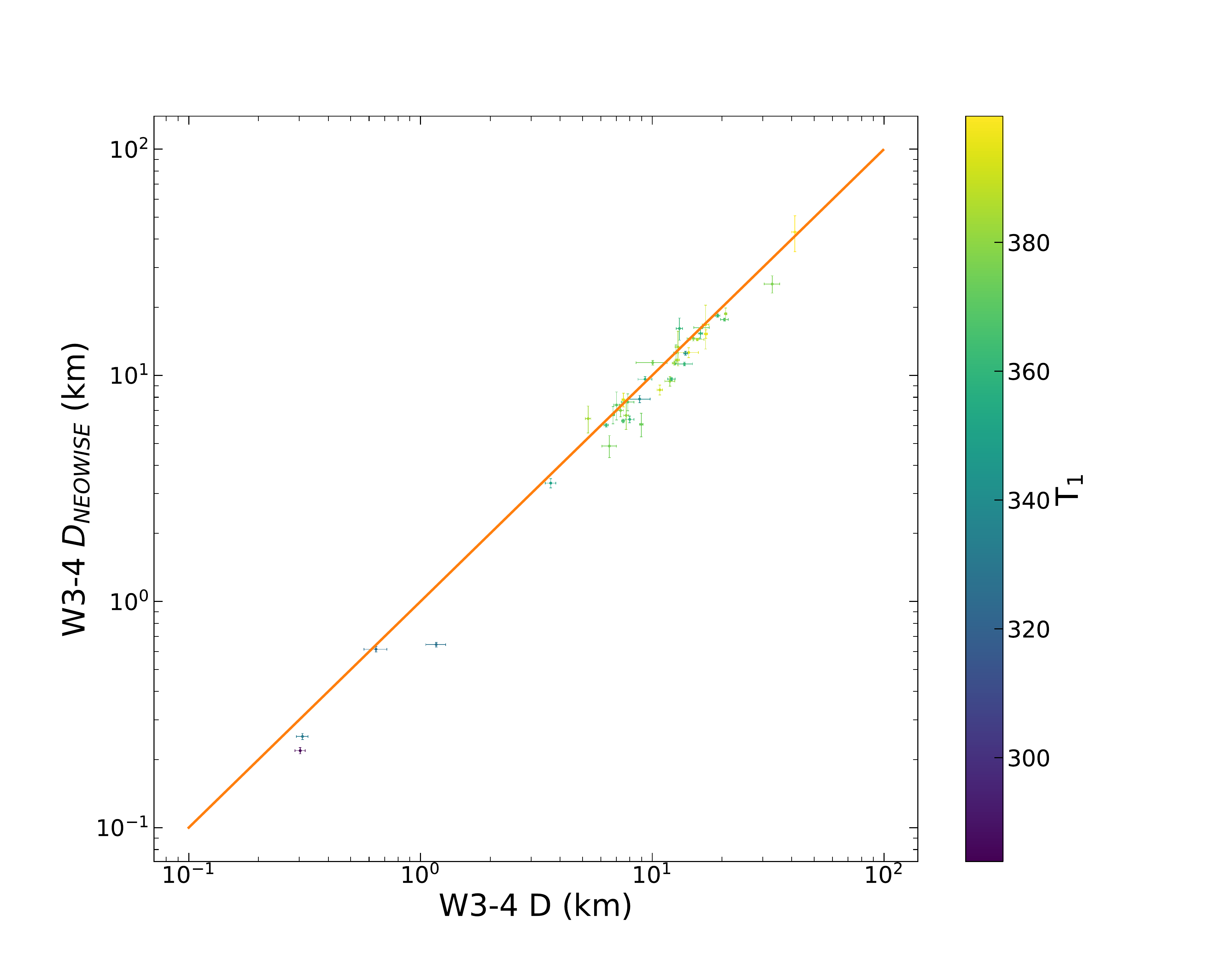}
\caption{Left: Plot of NEOWISE four-band diameters versus our corresponding diameters modeled with W3 and W4.  The orange line represents a perfect one-to-one ratio between diameters.  The color bar indicates our pseudo-temperature modeled with W3 and W4. Right: Plot of NEOWISE two-band (W3--4) diameters versus our corresponding diameters modeled with W3 and W4.}
\label{fig:neowise_w3w4}
\end{figure}

We also compared our two-band (W3--4) diameters to the diameters resulting from the four-band fits done by NEOWISE for the same set of asteroids.  We found 2651 asteroids in common, where 138 asteroids were represented in two different epochs, which therefore resulted in 2789 diameter comparisons plotted in Figure \ref{fig:neowise_w3w4}.  The median percent difference
$(D_{\rm W3-4} - D_{\rm W1-4,N})/D_{\rm W1-4,N}$ was 14\% (5\%--95\% quantiles at 2.0\% and 25\%).  
which suggests that the NEOWISE four-band results underestimate the diameter whereas our W3--4 results overestimate the diameter.  

For the two-band (W3--4) NEOWISE results, we found 42 asteroids in common and 0 asteroids with two different clusters of epochs, giving 42 total comparisons.  The median difference
$(D_{\rm W3-4} - D_{\rm W3-4,N})/D_{\rm W3-4,N}$
was 9.5\%
(5\%--95\% quantiles at -11\% and 38\%),
which suggests that the NEOWISE two-band results also underestimate the diameter while our W3--4 results overestimate the diameter.  

\section{Discussion}
The total flux from an object scales as $D^2$ (Equation \ref{eqn:F reflected Sun}) and the emitted component scales as $T^4$ (Stefan-Boltzmann Law).
When emitted light dominates, we expect $D$ to scale approximately with $T^2$.  Therefore, we expect a small difference in $T_1$ to produce a larger difference in $D$.  This trend is apparent in Figure \ref{fig:TD_reldiff}, where the median absolute value relative difference in pseudo-temperature is 2.42\%, whereas the median absolute value relative difference in diameter is larger, at 10.2\%.

We found that analysis of W2--3 observations tended to yield underestimates of asteroid diameters by about 15\% in comparison to occultation diameters, which we take to be the most accurate diameter measurements for this data set.  Such underestimates could potentially affect planetary defense efforts by underestimating the kinetic energy of potential impactors or by affecting Yarkovsky drift calculations.  This limitation affects analyses of existing W2--3 observations and likely of data from future missions with similar bands.

Additionally, as seen in Figures \ref{fig:TD_sigma} and \ref{fig:TD_sigma_w3w4}, we found a slightly larger median spread in bootstrap diameter estimates for the two-band fits (W2--3 and W3--4) than the four-band fits.  This larger spread indicates that two-band instruments may yield slightly larger uncertainties than analogous measurements obtained with four bands.

Our results are based on a curated subset of observations obtained during WISE's cryogenic mission phase.  This subset includes only asteroids that were detected in all four bands and may therefore exclude smaller or darker asteroids that were too dim to yield detections in W1--2.  Likewise, it excludes objects that resulted in saturated observations and may therefore exclude larger or brighter objects.  With these potential biases in mind, our sample covers a large range of diameters from 0.2 to 80 km as well as a large range of visible albedos.  As a result, it is expected to have broad applicability and relevance for future surveys.

We found that the two-band results for the near-Earth objects within our sample exhibit roughly the same characteristics as those for main belt objects.  However, our results required removal of anomalous low-temperature trial solutions, which
 were slightly more likely to originate
 from the NEO population due to their larger phase angles.  Our work suggests that future NEO surveys may benefit from the removal of unreasonable thermal model solutions for NEOs observed at large phase angles.

\label{sec-discussion}

\section{Conclusions}
In this study, we compared the results of the reparameterized Near-Earth Asteroid Thermal Model for a curated set of asteroids observed by WISE in two wavelength
bands (W2--3 and W3--4) to previous results from all four wavelength bands (W1--4) calculated by \citet{myhr22}.  This comparison was done with the goal
of elucidating unique aspects of determining asteroid parameters using modeling with two infrared bands so that any potential biases or shortcomings could be predicted
and corrected for with the upcoming NEO Surveyor mission, which will observe asteroids in two-bands similar to W2 and W3, or with similar future missions.

We found that the two-band W2--3 fits tended to have slightly higher pseudo-temperatures and smaller diameters than the four-band fits.  By comparison with the more accurate occultation diameters, we found that the W2--3 fits are likely
underestimating the size of the asteroids by about 15\%.  We also found that the W2--3 fits tend to have about the same median spread in pseudo-temperatures, but a slightly
larger spread in diameters across the bootstrapped fits, indicating a greater uncertainty in size of the asteroid as opposed to the four-band fits.  The W3--4 fits tended to give diameters off by about 12\% in comparison to the occultation results, and tended to have a larger diameter and lower pseudo-temperature than the corresponding four-band fits.  The W3--4 results show a larger median spread in diameters and pseudo-temperatures than the four-band fits, indicating greater uncertainty in size and pseudo-temperature than the four-band fits.  

Based on all of our results, we have concluded that
NEATM modeling of asteroids is worth doing even if there are data only in two of the four WISE wavelength bands, at least in the case of W2--3 and W3--4, because this modeling is still able to constrain asteroid sizes and albedos reasonably well.  Although the diameters appear to have a greater systematic error for W2--3 and W3--4 than for the four-band fits, these values are still valuable for understanding the asteroid population as a whole, and for being able to predict the relative danger posed by a potential impact event in a way that is more precise than using visible photometry alone.

\label{sec-conclusions}

EW was funded in part by the Nathan P. Myhrvold Graduate Fellowship.  AL was funded in part by the Joe and Andrea Straus Endowment for Undergrad Opportunity
and the Donald Carlisle Undergrad Research Endowed Fund.
We would like to thank two anonymous reviewers for helpful contributions that improved our work.

\software{
NumPy \citep{numpy},
SciPy \citep{scipy},
pandas \citep{pandas},
Matplotlib \citep{mpl}
}

\bibliographystyle{aasjournal}
\bibliography{wise_2band}

\begin{thebibliography}{}
\expandafter\ifx\csname natexlab\endcsname\relax\def\natexlab#1{#1}\fi
\providecommand{\url}[1]{\href{#1}{#1}}
\providecommand{\dodoi}[1]{doi:~\href{http://doi.org/#1}{\nolinkurl{#1}}}
\providecommand{\doeprint}[1]{\href{http://ascl.net/#1}{\nolinkurl{http://ascl.net/#1}}}
\providecommand{\doarXiv}[1]{\href{https://arxiv.org/abs/#1}{\nolinkurl{https://arxiv.org/abs/#1}}}

\bibitem[{Bowell {et~al.}(1989)Bowell, Hapke, Domingue, Lumme, Peltoniemi, \&
  Harris}]{bowell89}
Bowell, E., Hapke, B., Domingue, D., {et~al.} 1989, in Asteroids II, ed.
  R.~Binzel, T.~Gehrels, \& M.~Matthews (University of Arizona Press), 524--556

\bibitem[{Delbo(2004)}]{delbo04}
Delbo, M. 2004, PhD thesis, \dodoi{10.17169/REFUBIUM-14854}

\bibitem[{{Hanu{\v s}} {et~al.}(2015){Hanu{\v s}}, {Delbo'}, {{\v D}urech}, \&
  {Al{\'{\i}}-Lagoa}}]{hanu15}
{Hanu{\v s}}, J., {Delbo'}, M., {{\v D}urech}, J., \& {Al{\'{\i}}-Lagoa}, V.
  2015, Icarus, 256, 101, \dodoi{10.1016/j.icarus.2015.04.014}

\bibitem[{Harris(1998)}]{Harris_1998}
Harris, A.~W. 1998, Icarus, 131, 291, \dodoi{10.1006/icar.1997.5865}

\bibitem[{{Harris} \& {Lagerros}(2002)}]{Harris_and_Lagerros_2002}
{Harris}, A.~W., \& {Lagerros}, J. S.~V. 2002, in Asteroids {III} (University
  of Arizona Press), 205--218, \dodoi{10.2307/j.ctv1v7zdn4}

\bibitem[{Harris {et~al.}(2020)Harris, Millman, van~der Walt, Gommers,
  Virtanen, Cournapeau, Wieser, Taylor, Berg, Smith, Kern, Picus, Hoyer, van
  Kerkwijk, Brett, Haldane, Fernández~del Río, Wiebe, Peterson,
  Gérard-Marchant, Sheppard, Reddy, Weckesser, Abbasi, Gohlke, \&
  Oliphant}]{numpy}
Harris, C.~R., Millman, K.~J., van~der Walt, S.~J., {et~al.} 2020, Nature, 585,
  357–362, \dodoi{10.1038/s41586-020-2649-2}

\bibitem[{Herald {et~al.}(2019)Herald, Frappa, Gault, Hayamizu, Kerr, Moore, \&
  Giacchini}]{herald19}
Herald, D., Frappa, E., Gault, D., {et~al.} 2019, Small Bodies Occultations
  Bundle V3.0,  NASA Planetary Data System, \dodoi{10.26033/AP0G-WF63}

\bibitem[{Hunter(2007)}]{mpl}
Hunter, J.~D. 2007, Computing in Science Engineering, 9, 90,
  \dodoi{10.1109/MCSE.2007.55}

\bibitem[{Lam {et~al.}(2022)Lam, Margot, Whittaker, \& Myhrvold}]{lam23}
Lam, A. L.~H., Margot, J.-L., Whittaker, E., \& Myhrvold, N. 2022, Submitted to
  Planetary Science Journal, \dodoi{10.48550/ARXIV.2211.16409}

\bibitem[{Lebofsky {et~al.}(1986)}]{Lebofsky_1986}
Lebofsky, L.~A., {et~al.} 1986, Icarus, 68, 239,
  \dodoi{https://doi.org/10.1016/0019-1035(86)90021-7}

\bibitem[{{Mainzer} {et~al.}(2019){Mainzer}, {Bauer}, {Cutri}, {Grav},
  {Kramer}, {Masiero}, {Sonnett}, \& {Wright}}]{main19dps}
{Mainzer}, A.~K., {Bauer}, J.~M., {Cutri}, R.~M., {et~al.} 2019, NASA Planetary
  Data System, \dodoi{10.26033/18S3-2Z54}

\bibitem[{{Mainzer} \& {NEOCam Science Team}(2016)}]{Mainzer_2016b}
{Mainzer}, A.~K., \& {NEOCam Science Team}. 2016, in AAS/Division for Planetary
  Sciences Meeting Abstracts, Vol.~48, AAS/Division for Planetary Sciences
  Meeting Abstracts \#48, 327.01

\bibitem[{Mainzer {et~al.}(2016)}]{Mainzer_2016a}
Mainzer, A.~K., {et~al.} 2016, NASA Planetary Data System.
\newblock
  \url{https://pds.nasa.gov/ds-view/pds/viewDataset.jsp?dsid=EAR-A-COMPIL-5-NEOWISEDIAM-V1.0}

\bibitem[{{Masiero} {et~al.}(2011){Masiero}, {Mainzer}, {Grav}, {Bauer},
  {Cutri}, {Dailey}, {Eisenhardt}, {McMillan}, {Spahr}, {Skrutskie}, {Tholen},
  {Walker}, {Wright}, {DeBaun}, {Elsbury}, {Gautier}, {Gomillion}, \&
  {Wilkins}}]{masiero2011}
{Masiero}, J.~R., {Mainzer}, A.~K., {Grav}, T., {et~al.} 2011, \apj, 741, 68,
  \dodoi{10.1088/0004-637X/741/2/68}

\bibitem[{{Myhrvold}(2018{\natexlab{a}})}]{myhr18}
{Myhrvold}, N. 2018{\natexlab{a}}, \icarus, 303, 91,
  \dodoi{10.1016/j.icarus.2017.12.024}

\bibitem[{{Myhrvold}(2018{\natexlab{b}})}]{myhr18empirical}
---. 2018{\natexlab{b}}, \icarus, 314, 64, \dodoi{10.1016/j.icarus.2018.05.004}

\bibitem[{{Myhrvold} {et~al.}(2022){Myhrvold}, {Pinchuk}, \& {Margot}}]{myhr22}
{Myhrvold}, N., {Pinchuk}, P., \& {Margot}, J.-L. 2022, Planetary Science
  Journal, 3, 30, \dodoi{10.3847/PSJ/ac3232}

\bibitem[{Vereš {et~al.}(2015)Vereš, Jedicke, Fitzsimmons, Denneau, Granvik,
  Bolin, Chastel, Wainscoat, Burgett, Chambers, Flewelling, Kaiser, Magnier,
  Morgan, Price, Tonry, \& Waters}]{veres15}
Vereš, P., Jedicke, R., Fitzsimmons, A., {et~al.} 2015, Icarus, 261, 34,
  \dodoi{https://doi.org/10.1016/j.icarus.2015.08.007}

\bibitem[{{Virtanen} {et~al.}(2020){Virtanen}, {Gommers}, {Oliphant},
  {Haberland}, {Reddy}, {Cournapeau}, {Burovski}, {Peterson}, {Weckesser},
  {Bright}, {van der Walt}, {Brett}, {Wilson}, {Jarrod Millman}, {Mayorov},
  {Nelson}, {Jones}, {Kern}, {Larson}, {Carey}, {Polat}, {Feng}, {Moore}, {Vand
  erPlas}, {Laxalde}, {Perktold}, {Cimrman}, {Henriksen}, {Quintero}, {Harris},
  {Archibald}, {Ribeiro}, {Pedregosa}, {van Mulbregt}, \&
  {Contributors}}]{scipy}
{Virtanen}, P., {Gommers}, R., {Oliphant}, T.~E., {et~al.} 2020, Nature
  Methods, 17, 261, \dodoi{https://doi.org/10.1038/s41592-019-0686-2}

\bibitem[{{W}es {M}c{K}inney(2010)}]{pandas}
{W}es {M}c{K}inney. 2010, in {P}roceedings of the 9th {P}ython in {S}cience
  {C}onference, ed. {S}t\'efan van~der {W}alt \& {J}arrod {M}illman, 56 -- 61,
  \dodoi{10.25080/Majora-92bf1922-00a}

\bibitem[{{WISE Team}(2020)}]{All-sky20}
{WISE Team}. 2020, WISE All-Sky Single Exposure (L1b) Source Table,  IPAC,
  \dodoi{10.26131/IRSA139}

\bibitem[{{Wright} {et~al.}(2010){Wright}, {Eisenhardt}, {Mainzer}, {Ressler},
  {Cutri}, {Jarrett}, {Kirkpatrick}, {Padgett}, {McMillan}, {Skrutskie},
  {Stanford}, {Cohen}, {Walker}, {Mather}, {Leisawitz}, {Gautier}, {McLean},
  {Benford}, {Lonsdale}, {Blain}, {Mendez}, {Irace}, {Duval}, {Liu}, {Royer},
  {Heinrichsen}, {Howard}, {Shannon}, {Kendall}, {Walsh}, {Larsen}, {Cardon},
  {Schick}, {Schwalm}, {Abid}, {Fabinsky}, {Naes}, \& {Tsai}}]{wrig10}
{Wright}, E.~L., {Eisenhardt}, P. R.~M., {Mainzer}, A.~K., {et~al.} 2010, \aj,
  140, 1868, \dodoi{10.1088/0004-6256/140/6/1868}

\end{thebibliography}

\appendix{}
\renewcommand{\thefigure}{A\arabic{figure}}
\setcounter{figure}{0}

\section{Emissivity in Two-band Fits}
\label{app-eps}
\begin{figure}[h]
\centering\includegraphics[width=7cm]{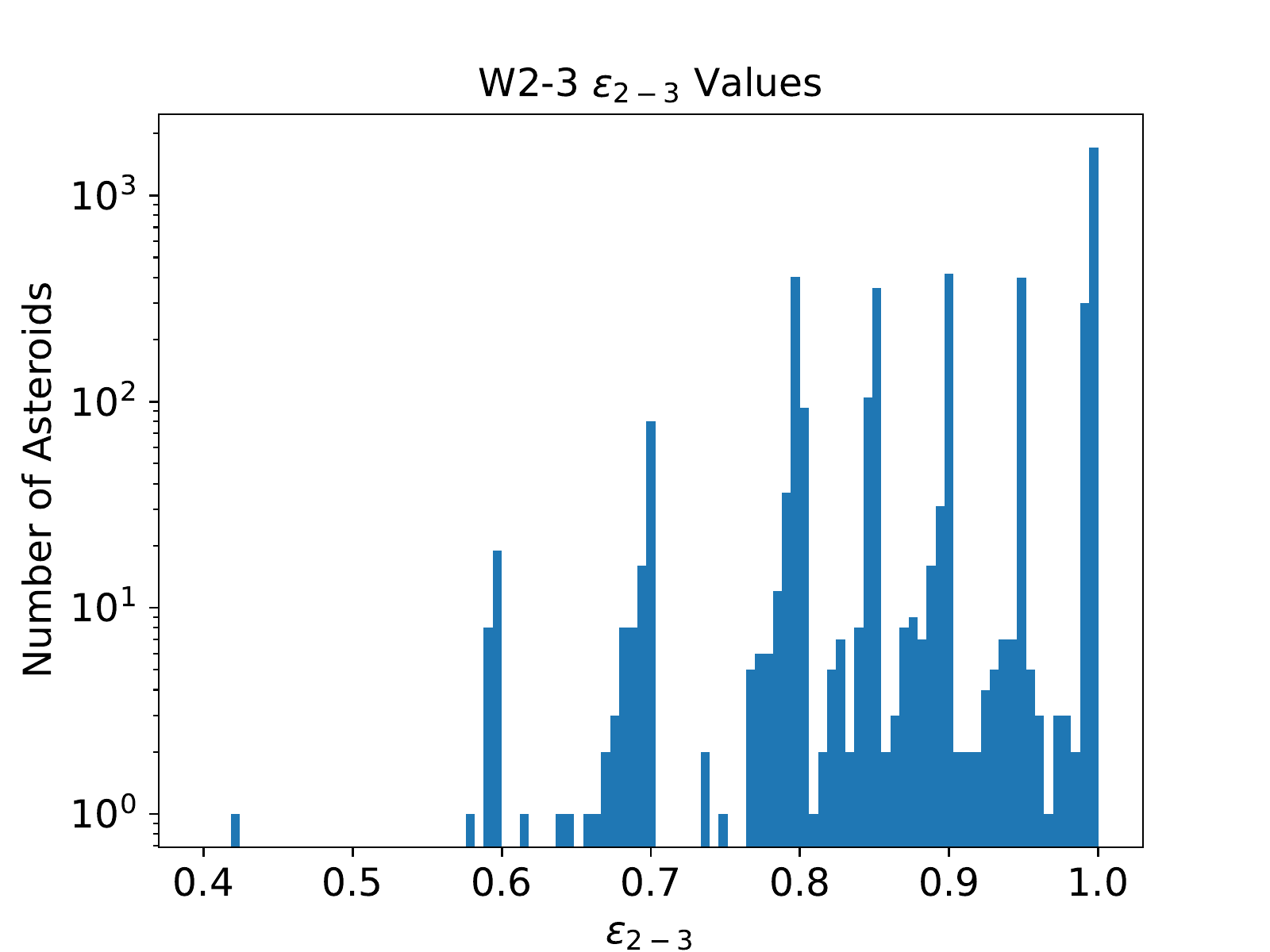}
\centering\includegraphics[width=7cm]{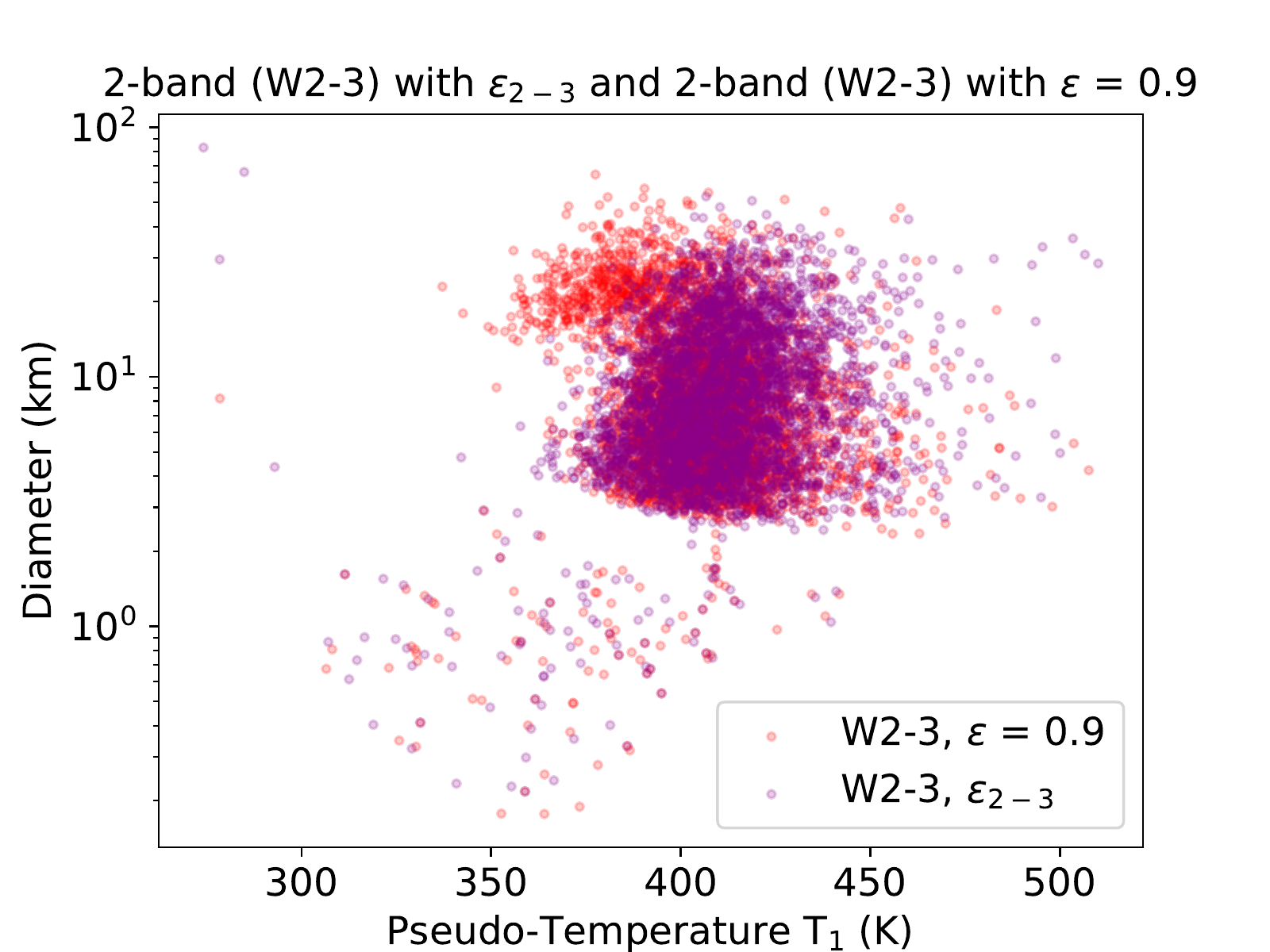}
\caption{(Left) Histogram of emissivity values obtained with the floating-emissivity model and W2--3.  (Right) Diameter vs.\ pseudo-temperature two-band estimates for the constant-emissivity model and the floating-emissivity model. }
\label{fig:combeps}
\end{figure}

To examine the validity of using a constant emissivity ($\epsilon=0.9$) in the two-band W2--3 fits,
we added one degree of freedom to our model and performed additional fits with a single floating value of the emissivity, $\epsilon_{2-3}$, which we applied to both bands.
We used the same minimization term with regularization that was used for four-band fits.  We ran a single model for each asteroid, using only the earlier cluster of observations
when two clusters were available.  We then eliminated asteroids for which either the constant-emissivity model or the floating-emissivity model gave a pseudo-temperature below 273 K, leaving a total of 4146 asteroids modeled.
The results appear to cover a range of emissivity values (Figure \ref{fig:combeps}, Left), but tend to cluster around the starting values of the emissivity trials, which included 0.6, 0.7, 0.8, 0.85, 0.9, 0.95, and 0.99.  This clustering indicates that the fitting algorithm
generally does not experience sufficient pressure to depart from the initial condition on emissivity, and that this parameter is therefore not essential.
The pseudo-temperature and diameter distributions are qualitatively similar (Figure \ref{fig:combeps}, Right).

We conducted an analysis of variance (also known as F-test) to determine whether the inclusion of the additional floating parameter ($\epsilon_{2-3}$) was statistically justified.
We found that the addition of the emissivity parameter cannot be deemed superfluous (p-value of 0.05) for only 5.5\% of our solutions.  In the other 94.5\% of solutions, the additional parameter is superfluous.
Because we expect about 5\% of cases to reach the critical p-value by definition,
the results of the F-tests strongly suggest that a fixed  emissivity is sufficient when fitting two infrared bands.

\section{Goodness of Fit Metric and Light Curve Variations}
\label{app-chisq}

\begin{figure}
\centering\includegraphics[width=8cm]{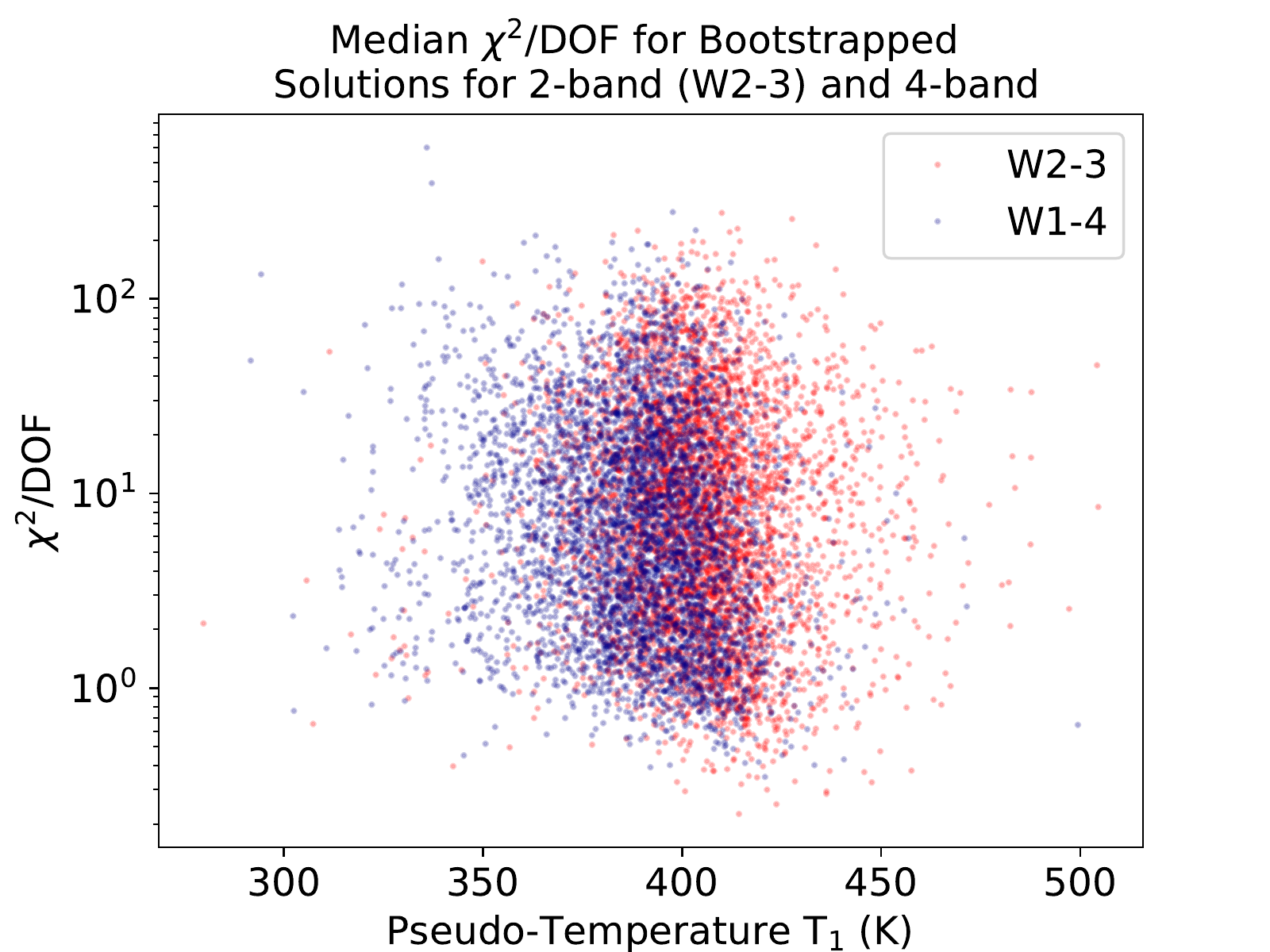}
\caption{Reduced chi-squared vs.\ pseudo-temperature.  Pseudo-temperatures and reduced chi-squared values are both represented by the median value of 200 bootstrapped trials.  Two-band values are shown in red whereas four-band values are shown in blue.}
\label{fig:chisqr_plot}
\end{figure}

Although we did not include WISE flux uncertainties when minimizing the SSR, we were interested in evaluating the traditional reduced chi-squared value as an indicator of goodness of fit.  For each cluster of observations, we calculated the median value over 200 bootstrap trials of 
\begin{equation}
\chi^2_\nu = \frac{\chi^{2}}{{\rm DOF}} = \frac{1}{N-p} \sum_{i = 1}^N \frac{(O_i - C_i)^{2}}{\sigma_i^2},
\label{eqn:chisqr}
\end{equation}
where $N$ is the number of data points, $p$ is the number of parameters being fit by the model, $O_i$ is an observed flux, $C_i$ is the corresponding modeled (computed) flux, and $\sigma_i$ is the flux uncertainty (Figure \ref{fig:chisqr_plot}).  The instances of the four-band bootstrap trials used to calculate $\chi^2_\nu$, the reduced chi-squared, are different from those of \citet{myhr22}, but we expect essentially the same results.

The WISE flux uncertainties
have been found in previous research by \citet{hanu15} and \citet{myhr18empirical} 
to be
underestimated.  To account for this problem, \citet{myhr22} used ``double detections'' of thousands of objects to derive 
more accurate uncertainties.  They found correction factors that can be multiplied by the WISE uncertainties to obtain more realistic uncertainty values. 
These correction factors are 1.224 for W1, 1.120 for W2, 1.479 for W3, and 1.218 for W4 \citep{lam23}, and they have been applied to the $\sigma_i$ used in equation (\ref{eqn:chisqr}).

\begin{figure}
\centering\includegraphics[width=8cm]{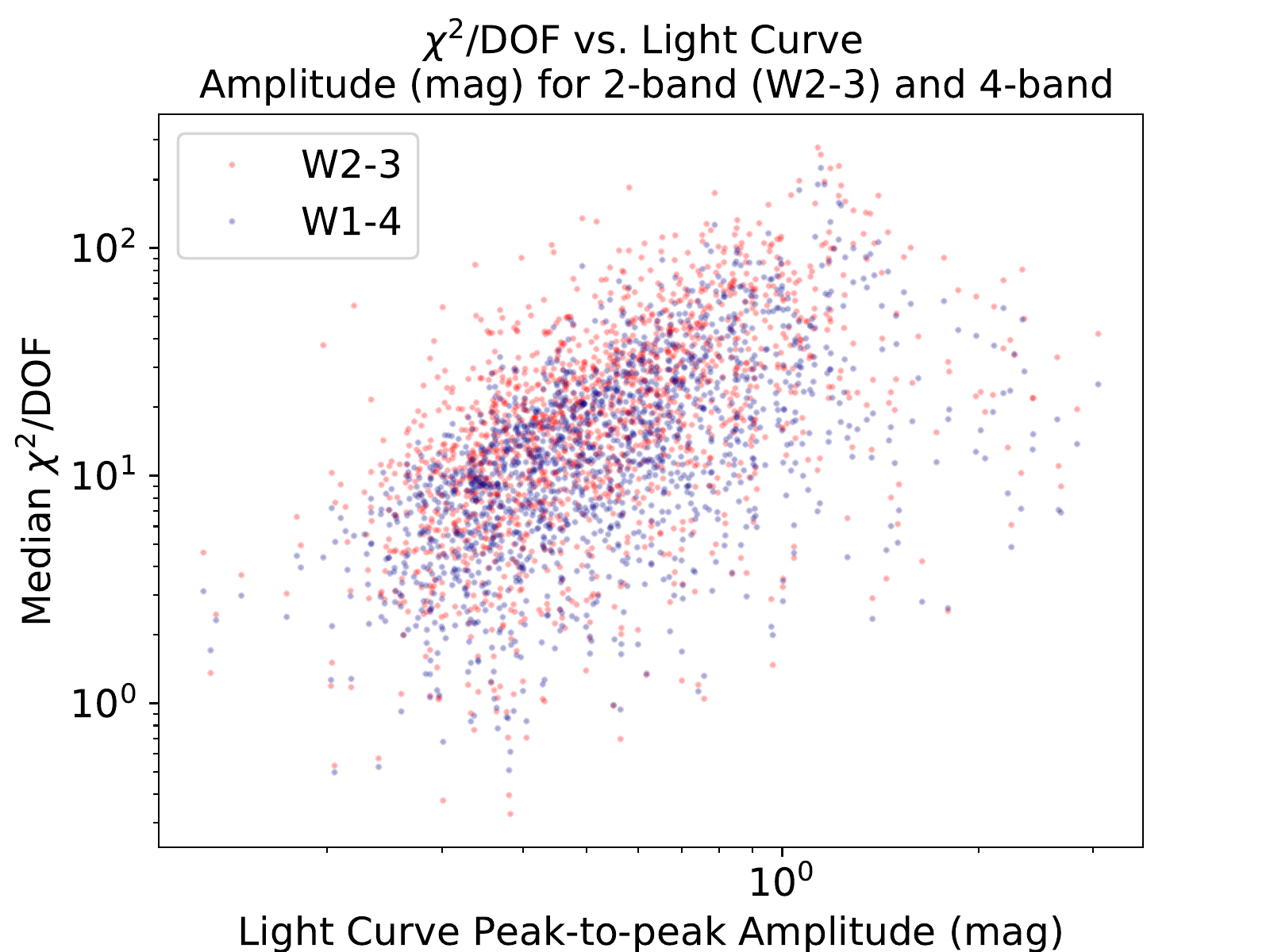}
\caption{Logarithmic plot of lightcurve peak-to-peak amplitude vs.\ reduced chi-squared for all remaining clusters with a corresponding lightcurve fit.  W2--3 clusters are shown in red whereas W1--4 clusters are shown in blue.  Lightcurve amplitudes were obtained from \citet{lam23}.}
\label{fig:lc_amp_redchisqr}
\end{figure}

The median values of reduced chi-squared are larger than unity ($\chi_\nu^2$ = 7.6 for W2--3 and $\chi_\nu^{2}$ = 6.4 for W1--4), and some reduced chi-squared values exceed 100.  We ascribe the large reduced chi-squared values primarily to lightcurve variations, which are not accounted for in our model.
Reduced chi-squared values are indeed correlated with light curve amplitude, with a correlation coefficient of 0.48 between the logarithm of the amplitudes and the logarithm of the two-band reduced chi-squared values and a correlation coefficient of 0.50 between the logarithm of the amplitudes and the logarithm of the four-band reduced chi-squared values. (Figure \ref{fig:lc_amp_redchisqr}).

\begin{figure}
\centering\includegraphics[width=14cm]{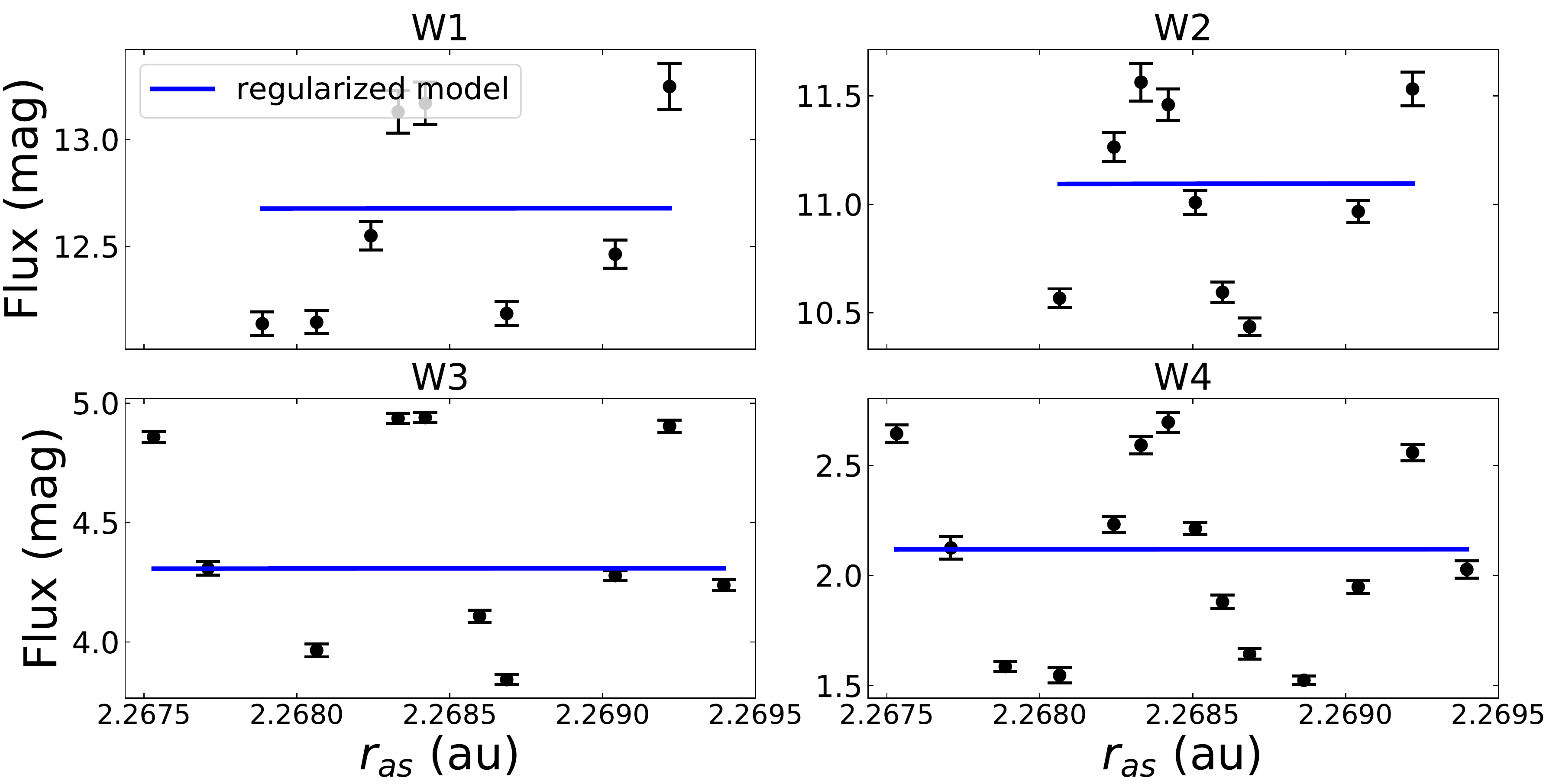}
\caption{W1--4 flux values vs.\ asteroid-Sun distance for asteroid 1182.  Black points and error bars indicate the WISE fluxes and uncertainties, whereas blue
lines indicate the best four-band fit to the data.  Error bars incorporate the correction factors.}
\label{fig:1182}
\end{figure}

The observations of asteroid 1182 (Figure~\ref{fig:1182}) provide an example of this phenomenon.  This asteroid has a light curve amplitude of 1.16 mag and a reduced chi-squared of 196 for the W2--3 model and 191 for the W1--4 model.

\section{Anomalous Low-Temperature Solutions and Distance to the Sun}
\label{app-lowT}
\begin{figure}
\centering\includegraphics[width=8cm]{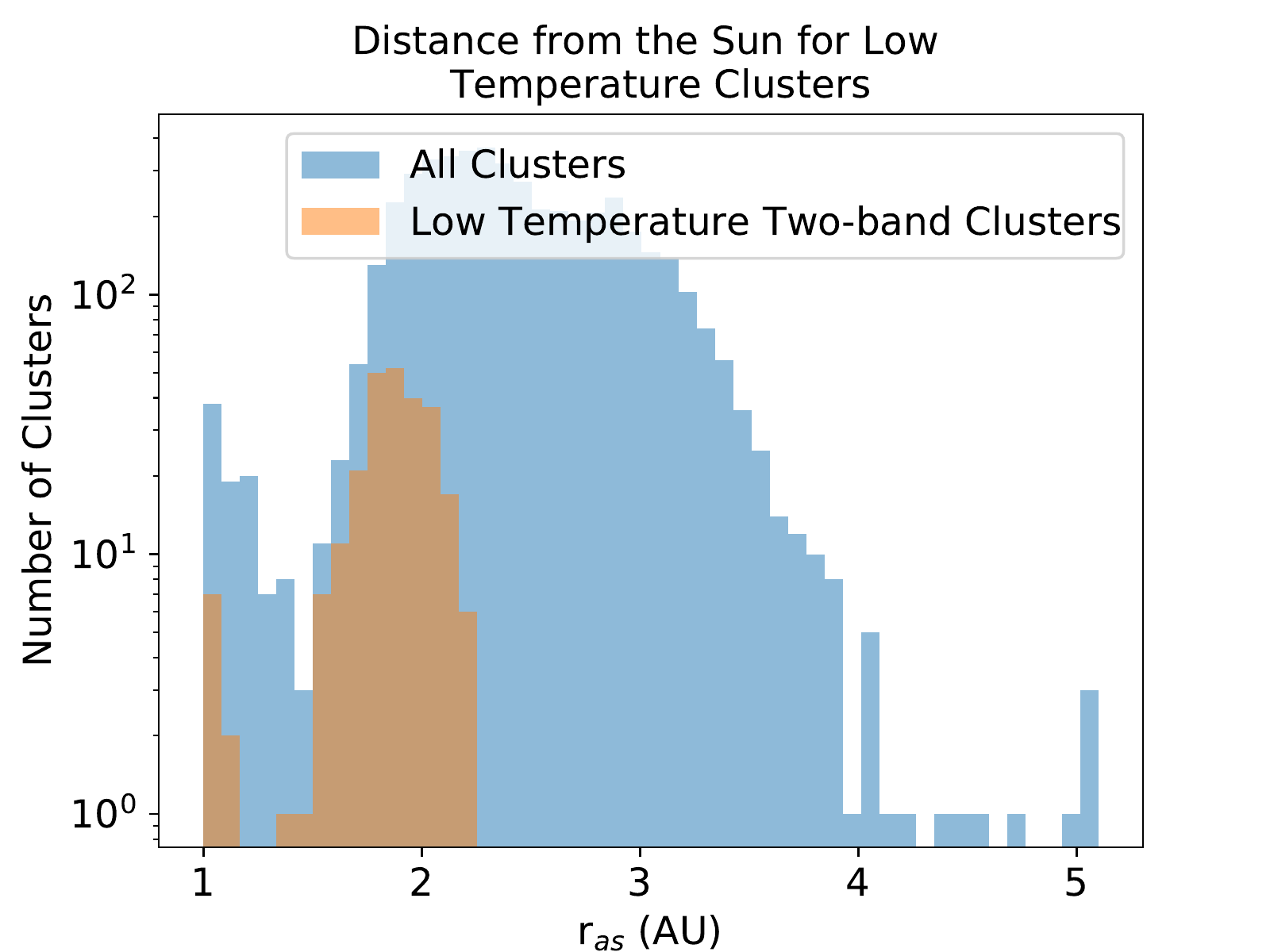}
\centering\includegraphics[width=8cm]{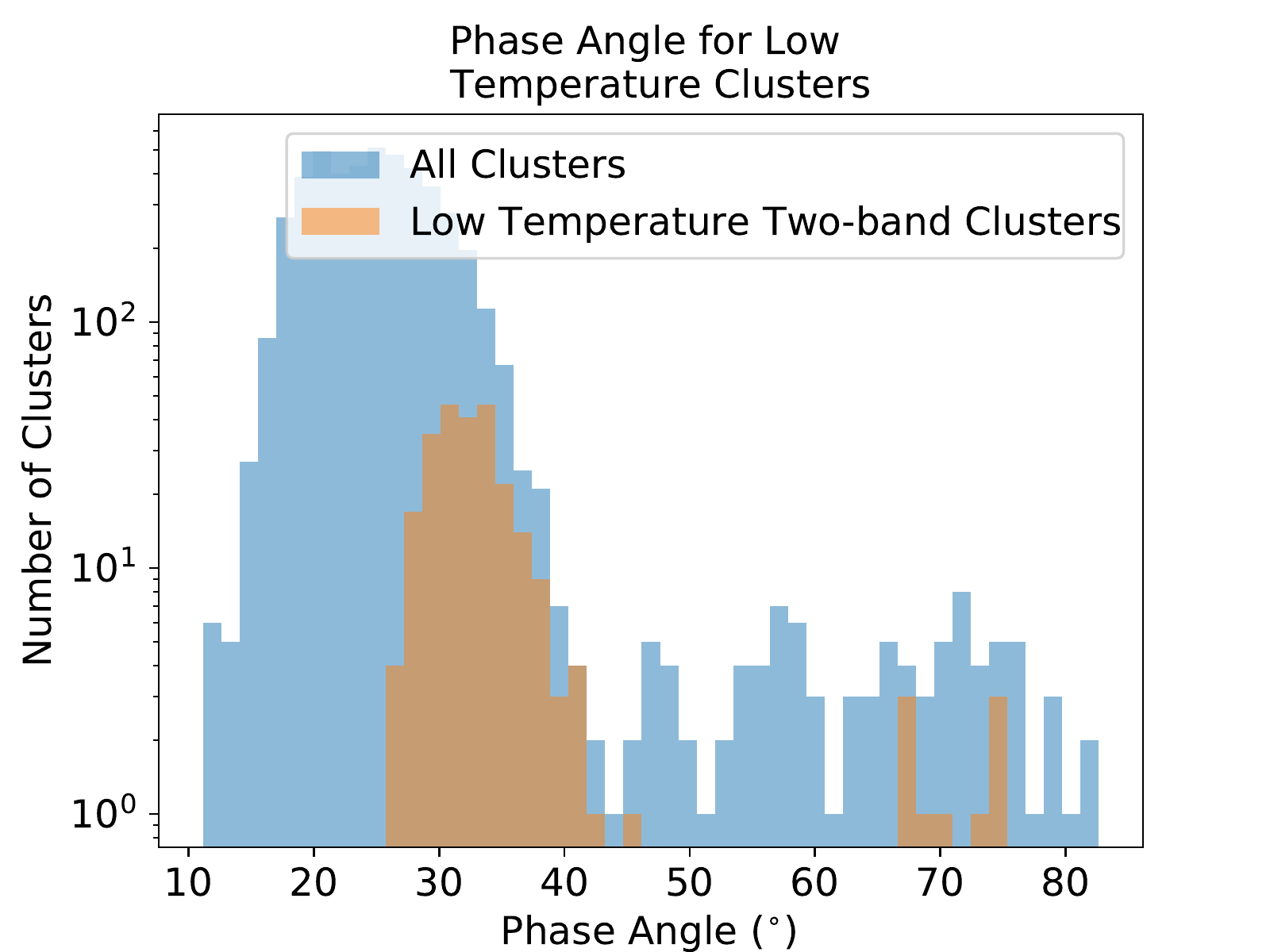}
\caption{Histograms of distance from the Sun (Left) and phase angle (Right) at the time of observations.  The blue bars represent all observational clusters, while the orange bars represent clusters that had a low median pseudo-temperature (T$_1$ $<$ 273 K) from the two-band bootstrapped fits, before implementation of our low-temperature solution removal process. }
\label{fig:ras}
\end{figure}

In our exploration of the factors contributing to low pseudo-temperature values in some W2--3 solutions, we found a correlation between the
asteroid-Sun distance
during observations
and the median pseudo-temperature of that cluster from the bootstrapped trials, as well as a correlation between phase angle and median pseudo-temperature (Figure \ref{fig:ras}).
With the WISE observing geometry, asteroids observed at small distances from the Sun generally have larger phase angles than asteroids located further away.
We suspect that anomalous thermal solutions are more likely for asteroids at large phase angles due to NEATM limitations, errors in the assumed slope parameter ($G$) value, and the fact that the HG phase function \citep[][but see correction in \citet{myhr22}]{bowell89} does not adequately account for the true brightness expected at larger phase angles.  
We found that 4.8\% of the W2--3 low pseudo-temperature clusters represent near-Earth objects, which is somewhat higher than the 2.0\% of all W2--3 clusters that are near-Earth objects.
  Most of the low-temperature solutions are due to asteroids with small asteroid-Sun distances and relatively large phase angles, such as inner main belt objects or Mars-crossing objects.

\section{Robustness with Respect to Minimization Technique}
\label{app-2vs4}

\begin{figure}
\centering\includegraphics[width=5cm]{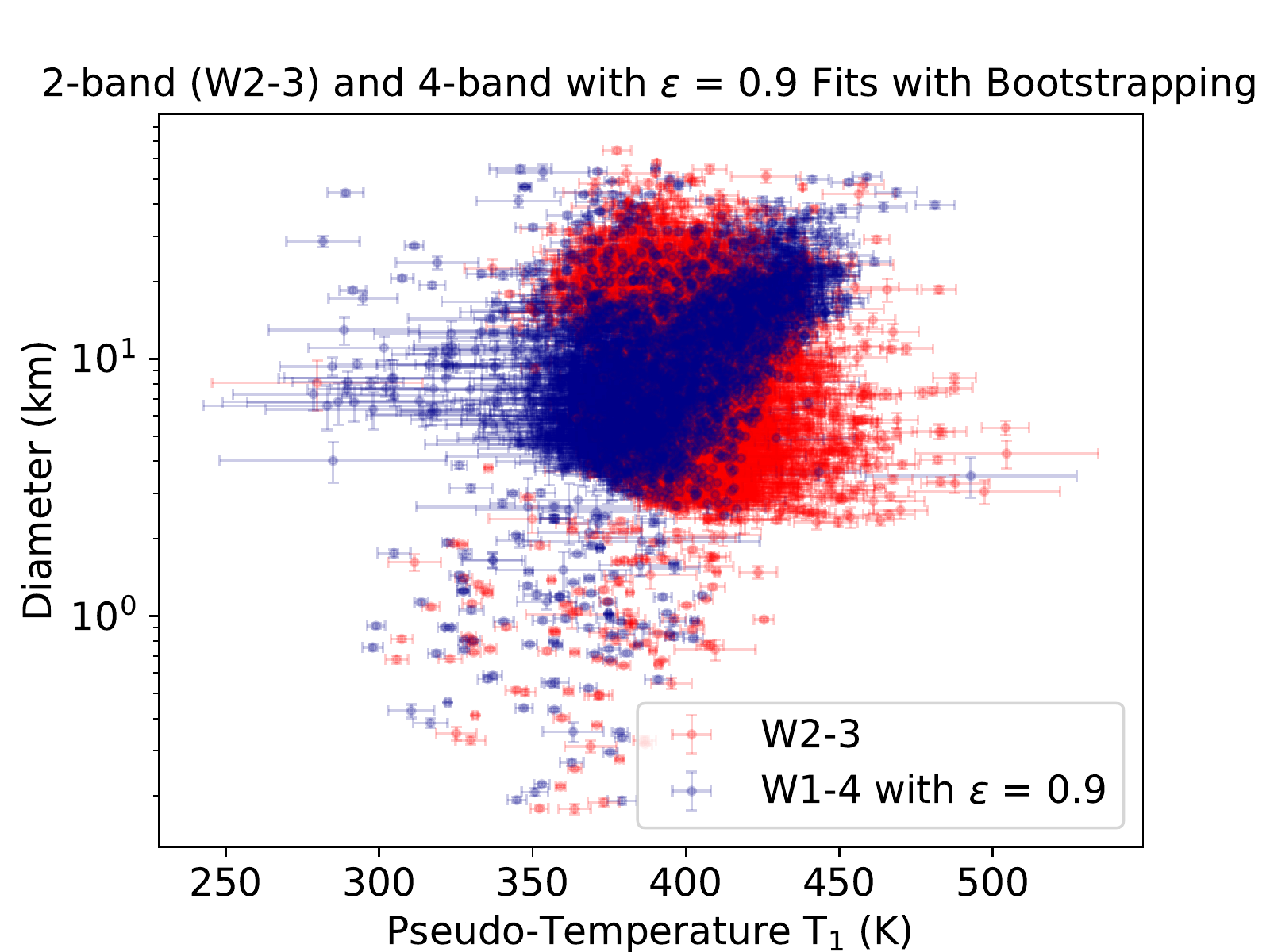}
\centering\includegraphics[width=5cm]{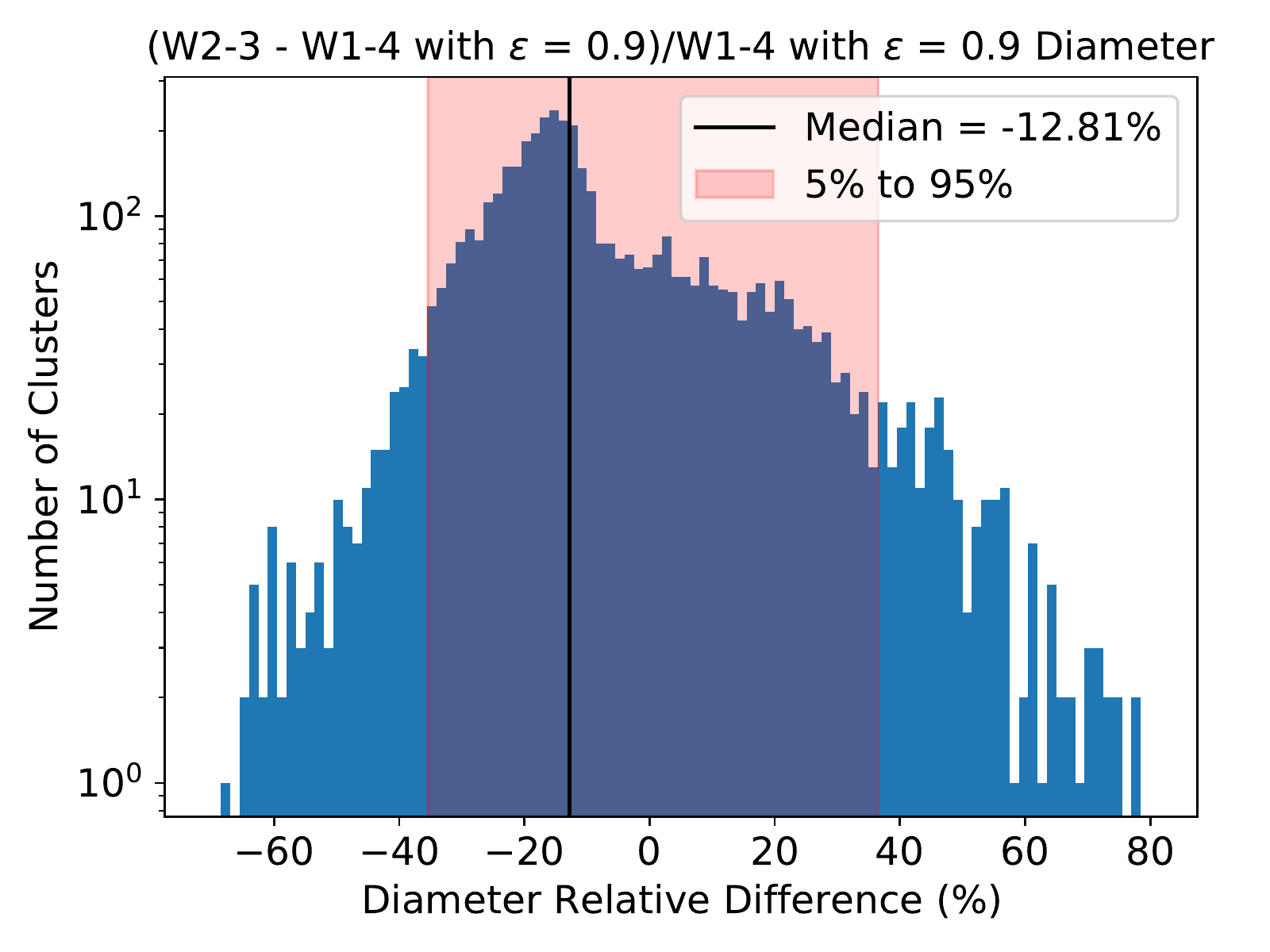}
\centering\includegraphics[width=5cm]{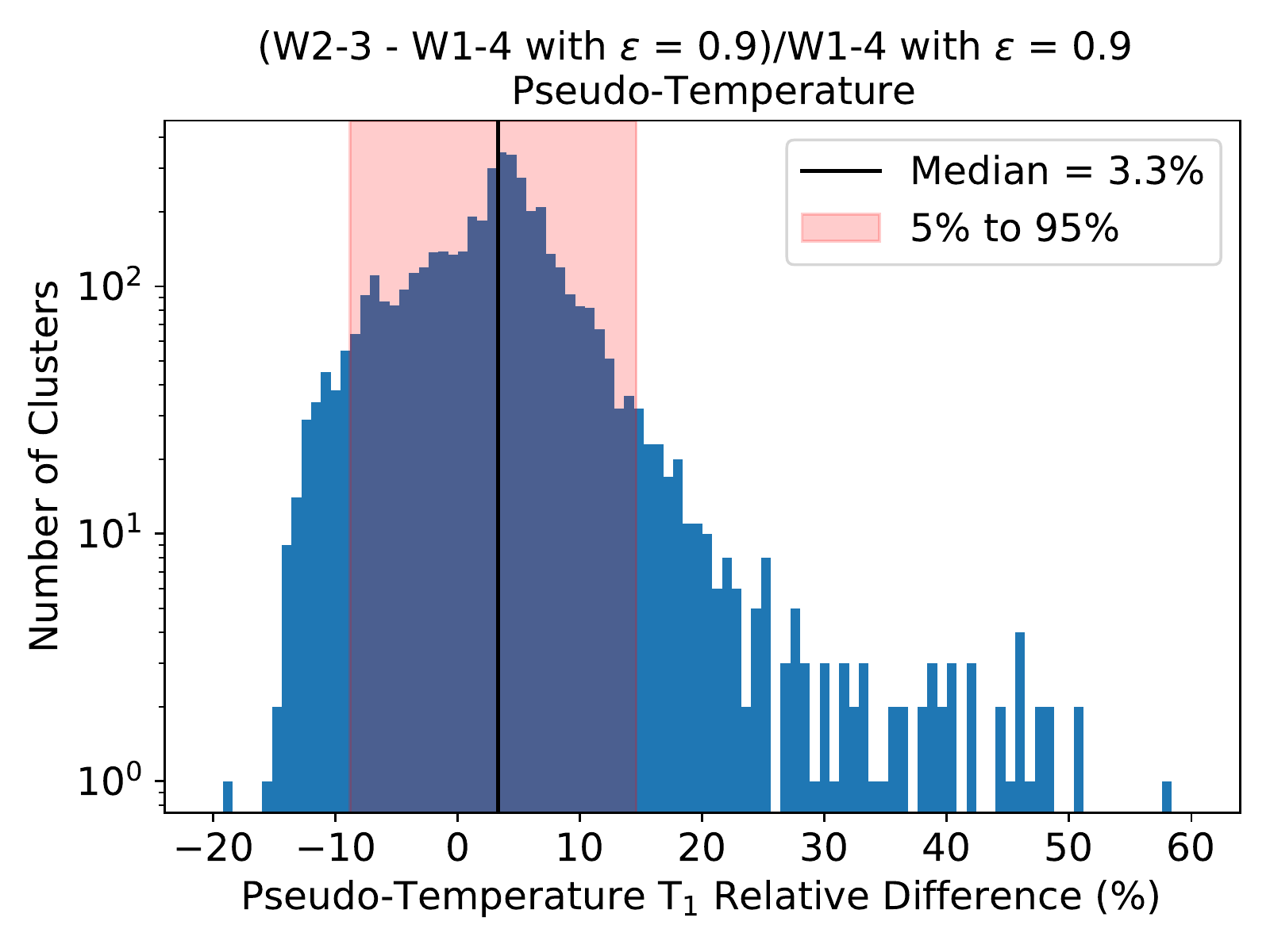}
\caption{Comparison of results from two-band W2--3 fits to results from four-band fits with $\epsilon$ fixed to 0.9.  Left: Diameter vs.\ pseudo-temperature $T_1$ for 4533 clusters.  Each point represents the median diameter and pseudo-temperature value among all accepted bootstrap trials for an object while error bars represent the standard deviations of the diameter and pseudo-temperature values among accepted bootstrap trials.  Middle:  Histogram of the relative difference in diameter.  Median values are represented with a vertical black line while a pink highlighted block is used to represent the 5 - 95\% quantile range.  Right: Histogram of relative difference in pseudo-temperature.}
\label{fig:4band_eps_pt9}
\end{figure}

In addition to the nominal four-band fits, we also conducted four-band fits where we fixed the emissivity to 0.9 and minimized the SSR, in order to replicate the two-band fits as closely as possible.  The only difference between the two-band fits and these fixed-emissivity four-band fits is the number of wavelength bands used in the fitting process.
The distribution of W1--4 solutions with fixed emissivity (Figure \ref{fig:4band_eps_pt9}) does look significantly different from the distribution of nominal four-band fit results (Figure \ref{fig:TD_bootstrap}).  However, the difference in median diameter between the two-band and four-band fits remains very similar, $(D_{\rm W2-3} - D_{\rm W1-4,\epsilon=0.9})/D_{\rm W1-4,\epsilon=0.9}$, = -13\%, compared to the -10\% observed with the nominal four-band fits. 
The difference in median pseudo-temperatures is also similar, 3.3\% vs.\ 2.4\% with the nominal four-band fits.  This consistency in results suggests that the median offsets in pseudo-temperature and diameter observed between the two-band and four-band fits is due to the consequences of eliminating W1 and W4 from the fitting procedure rather than any difference in fitting methods between the two-band and four-band fits.  

\section{Systematic Error Introduced from Constant Emissivity}
\label{app-4vs4pt9}

\begin{figure}
\centering\includegraphics[width=5cm]{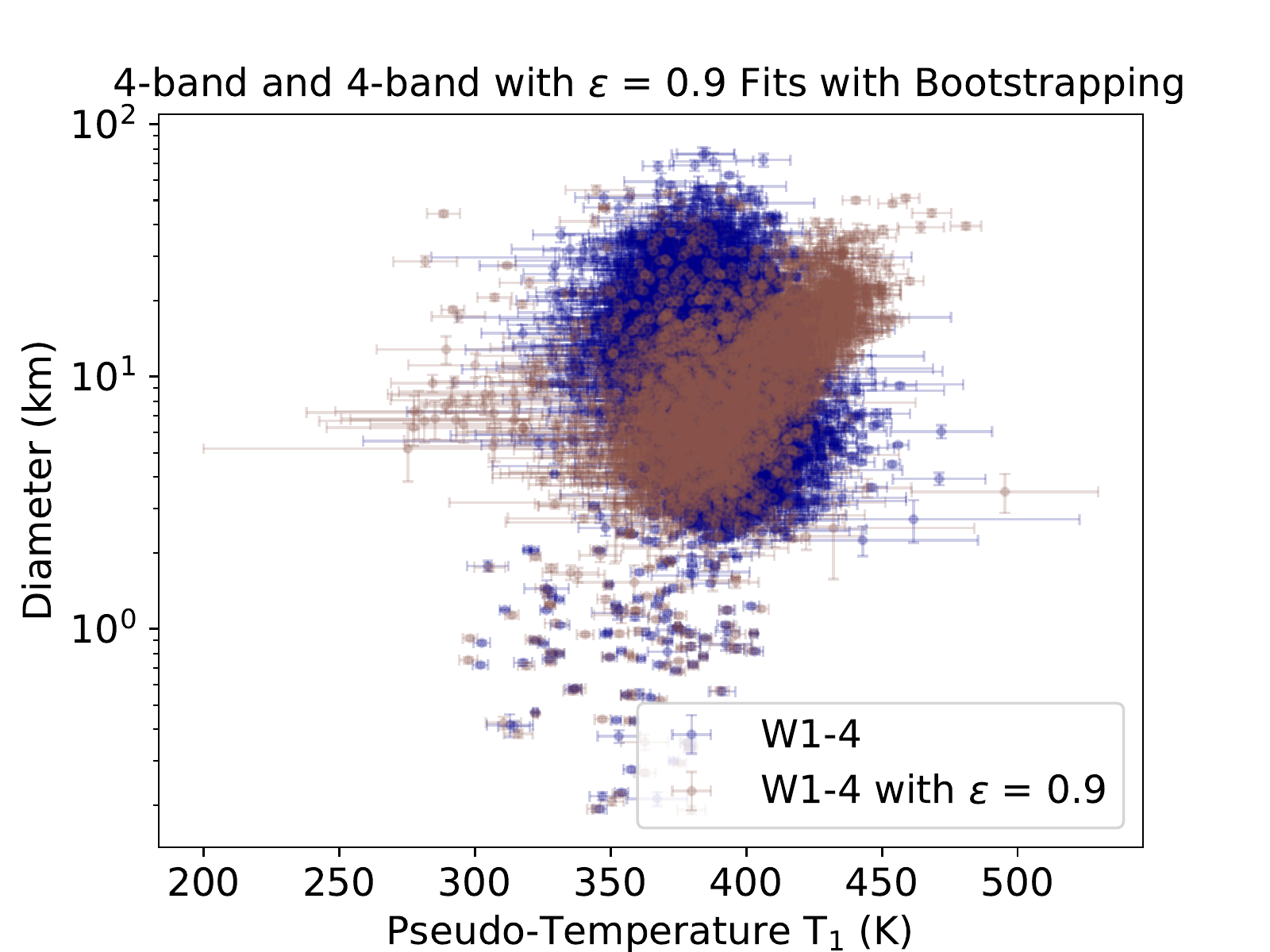}
\centering\includegraphics[width=5cm]{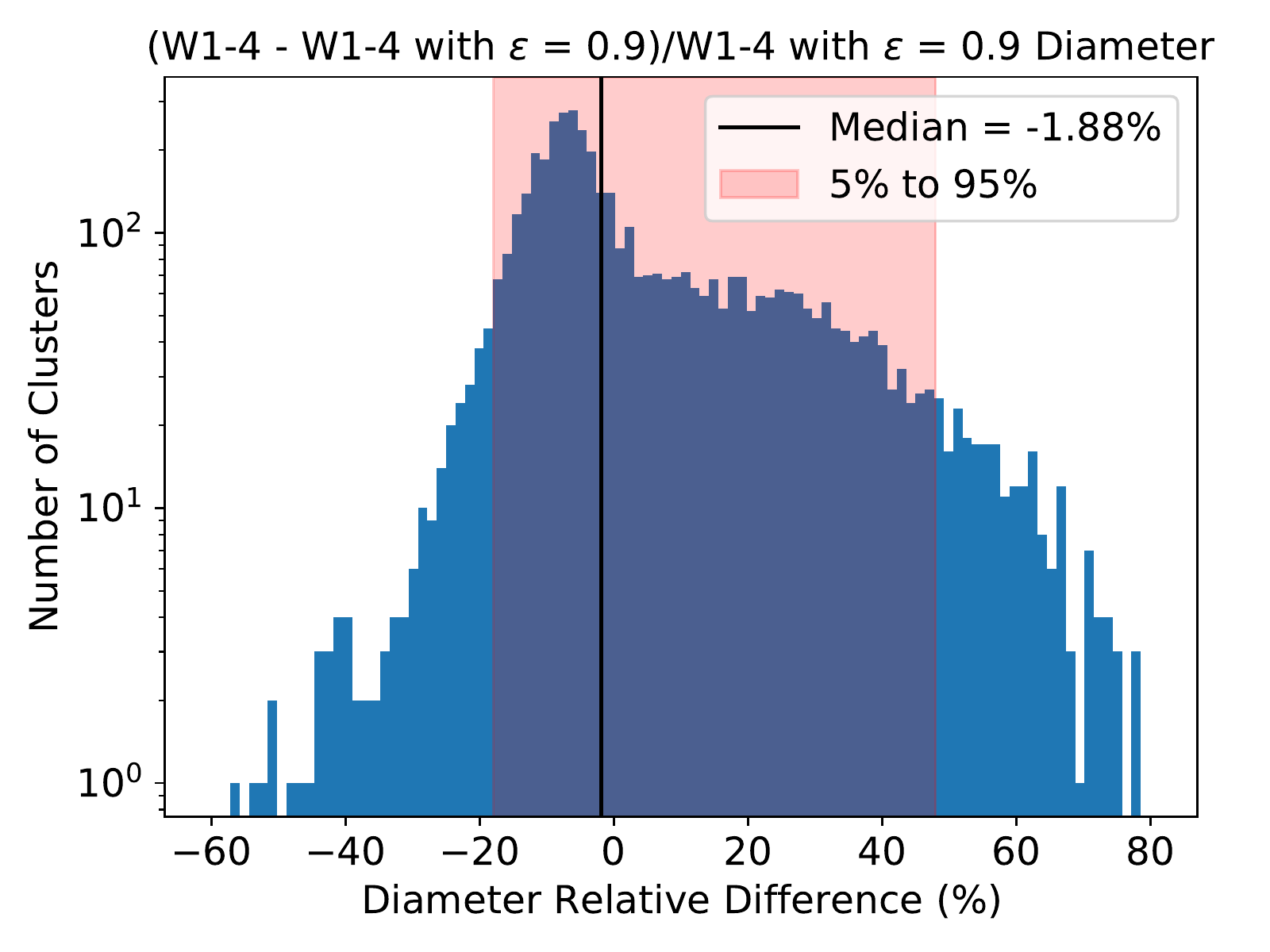}
\centering\includegraphics[width=5cm]{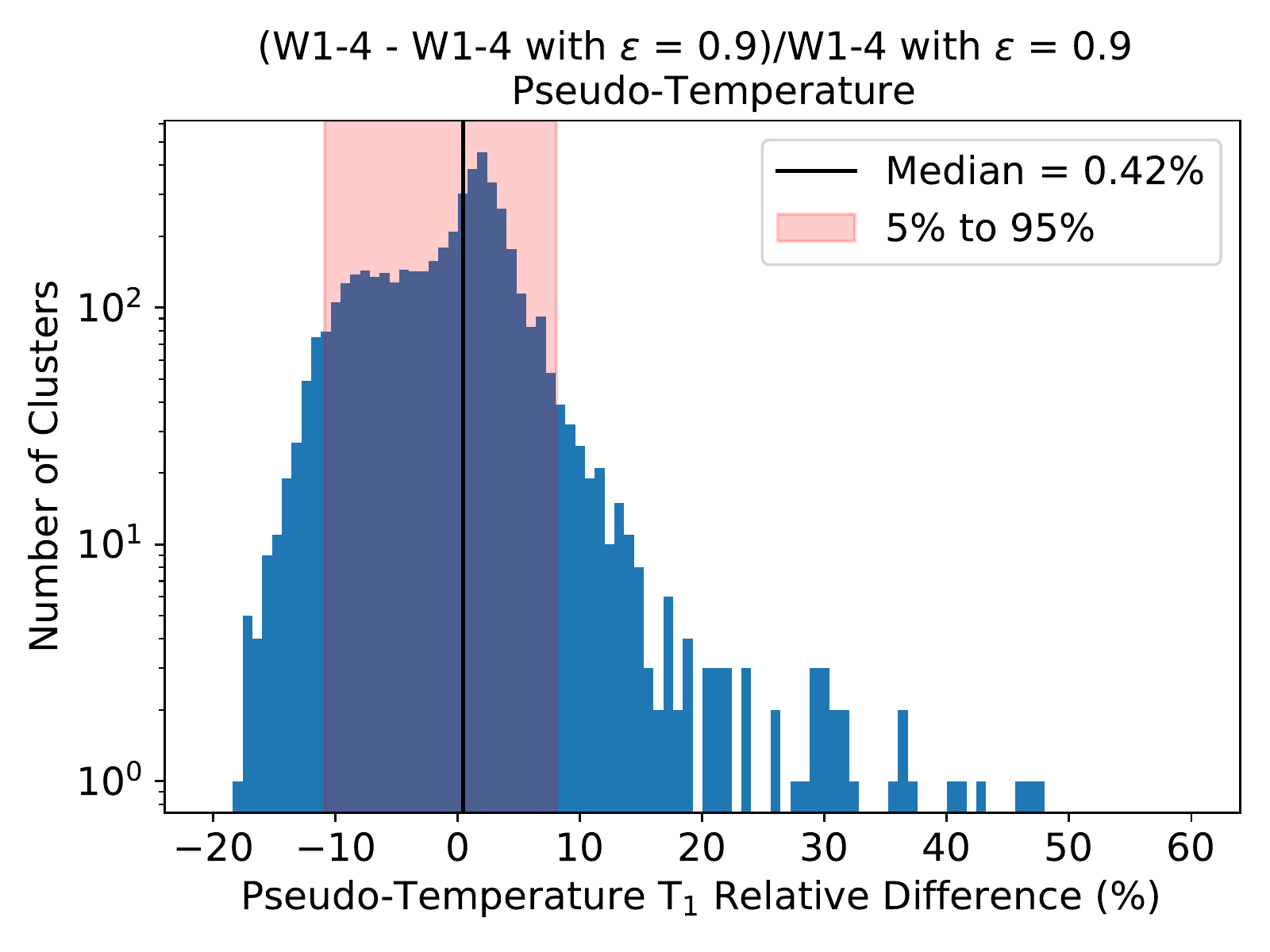}
\caption{Comparison of results from \citet{myhr22} four-band fits to results from four-band fits with $\epsilon$ fixed to 0.9.  Left: Diameter vs.\ pseudo-temperature $T_1$ for 4672 clusters.  Each point represents the median diameter and pseudo-temperature value among all accepted bootstrap trials for an object while error bars represent the standard deviations of the diameter and pseudo-temperature values among accepted bootstrap trials.  Middle:  Histogram of the relative differences in diameter.  Median values are represented with a vertical black line and the 5 - 95\% quantile range is represented with a region highlighted in pink.  Right: Histogram of relative differences in pseudo-temperature.}
\label{fig:4band_v_4band_eps_pt9}
\end{figure}

Using a fixed emissivity at 0.9 in thermal models introduces systematic errors.  To quantify this error, we compared results of the \citet{myhr22} four-band fits to four-band fits obtained with an emissivity fixed to 0.9 and with an SSR minimization term rather than a regularized minimization term.  The diameter and pseudo-temperature estimates are noticeably different (Figure \ref{fig:4band_v_4band_eps_pt9}).  The median diameter difference $(D_{\rm W1-4} - D_{\rm W1-4,\epsilon=0.9})/D_{\rm W1-4,\epsilon=0.9}$ is deceptively small at -1.88\%, but the 5\% and 95\% quantiles of the distribution show diameter differences of -18\% to 48\%.

\end{document}